\documentclass[aps,10pt,notitlepage,nofootinbib,twocolumn]{revtex4-2}
\usepackage[utf8]{inputenc}
\usepackage{amsmath,amssymb}
\usepackage{newtxtext,newtxmath}
\usepackage{appendix}
\usepackage{bbold}
\usepackage{bm}
\usepackage{graphicx}
\usepackage[usenames,dvipsnames]{xcolor}
\usepackage{color}
\usepackage{xcolor}
\usepackage{cuted}
\usepackage{slashed}
\usepackage[english]{babel}
\usepackage{dcolumn}
\usepackage{pifont}
\usepackage{dsfont,mathrsfs}
\usepackage{cancel}
\usepackage{bigints}
\usepackage{lineno}
\usepackage[colorlinks=true,linktocpage=true,linkcolor=blue,citecolor=blue]{hyperref}
\usepackage{cleveref}
\usepackage{booktabs} 
\usepackage{array}    

\usepackage{amsfonts}
\usepackage{float}
\usepackage{subfigure}
\usepackage{mathtools,slashed}
\usepackage{ulem}
\def\be {\begin{equation}}
	\def\ee {\end{equation}}
\def\bea {\begin{eqnarray}}
	\def\eea {\end{eqnarray}}
\def\bc {\begin{center}}
	\def\ec {\end{center}}

\def\mn {\mu\nu}

\def\nn {\nonumber}

\allowdisplaybreaks
\begin{document}
	
	
	\title{Shear and bulk viscosity of quark-gluon plasma with Gribov gluons and quasiparticle quarks}

	\author{Sadaf Madni$^{a}$}
	\email{ sadaf.madni@niser.ac.in}
	\author{Arghya Mukherjee$^{b}$}
	\email{arbp.phy@gmail.com}
	\author{Amaresh Jaiswal$^{a}$}
	\email{a.jaiswal@niser.ac.in}
	\author{Najmul Haque$^{a}$}
	\email{nhaque@niser.ac.in}

	\affiliation{$^a$School of Physical Sciences, National Institute of Science Education and Research, An OCC of Homi Bhabha National Institute, Jatni-752050, India}
	\affiliation{$^b$Ramakrishna Mission Residential College (Autonomous), Narendrapur, Kolkata-700103, India  }	
	
\begin{abstract}
In this study, we analyze the transport properties of the Quark-Gluon Plasma, focusing on bulk ($\zeta$) and shear ($\eta$) viscosities at vanishing chemical potential. To describe the QGP, we employ a quasiparticle model for quarks along with Gribov’s prescription for gluons, which effectively captures non-perturbative dynamics. The Gribov parameter $\gamma_G$ and the dynamical mass $m_g$ are obtained by solving the one-loop gap equation in the $\overline{\text{MS}}$ renormalization scheme and further using lattice QCD data for the equation of state (EoS) of pure gluonic matter. The interaction between quarks and gluons is reflected in the quark quasi-mass $m_q$, again obtained using lattice EoS data for (2+1)-flavor QCD. Our primary goal is to invertigate the influence of quasi-quarks on the transport coefficients of QGP. Interestingly, we find a substantial decrease in the scaled transport coefficients with rising temperatures within the range ($1 \le T/T_c \le 3.5$).
\end{abstract}
	
\maketitle
\section{Introduction}
The Large Hadron Collider (LHC) at CERN, Geneva, and the Relativistic Heavy Ion Collider (RHIC) at the Brookhaven National Laboratory (BNL) in New York are two remarkable scientific facilities that serve as vital platforms for studying the primordial form of matter that existed in the universe few microseconds after the Big Bang~\cite{Chen:2014kpa}. High-energy nucleus-nucleus collision experiments at these facilities have the potential to convert a portion of the kinetic energies of the colliding nuclei into heating the QCD vacuum within an extremely small volume. This could result in temperatures millions of times hotter than the core of the sun. At temperatures around $\sim$150 MeV, the transition of nuclear matter from the hadronic phase, consisting of protons and neutrons, to the deconfined phase, consisting of quarks and gluons, occurs. This transition from the hadronic phase to the deconfined phase leads to the emergence of a new state of matter termed the Quark-Gluon Plasma (QGP)~\cite{Pasechnik:2016wkt}. 
	
Analysis of spectra and azimuthal anisotropy of particles produced in relativistic heavy-ion collisions suggested that the QGP produced is strongly coupled~\cite{Braun-Munzinger:2007edi, Gyulassy:2004zy, Shuryak:2008eq, Heinz:2005zg}. Application of viscous hydrodynamics to understand the evolution of the fireball produced in heavy-ion collisions gained widespread interest ever since the ratio of shear viscosity to entropy density $\eta/s$ was estimated to be close to the absolute lower bound for all fluids, which  is $\hbar/(4\pi k_B)$~\cite{Kovtun:2004de}. This allows one to investigate the behaviour of QGP in the light of viscous hydrodynamics~\cite{Jaiswal:2016hex, Arnold:2012dqa, Florkowski:2015lra, Florkowski:2010zz}. Apart from the equation of state, transport coefficients such as shear and bulk viscosities are major supplements in the hydrodynamic equation, which governs the evolution of the fireball. While the equation of state for QGP is well settled from lattice QCD calculations, the transport coefficients are still not very well conclusive, particularly near the phase transition regime where huge uncertainity is observed due to the medium complexities. Therefore, a consistant formulation of the transport coefficients are still an open challenge and still awaits a final nail to it. 
	
Considering these, there has been major efforts in the past to determine the ratio $\eta/s$ as well as the ratio of bulk viscosity to entropy density $\zeta/s$ of the QGP as a function of the temperature $T$. The transport coefficients of QGP have been an active area of research for more than two decades and have been dealt with various approaches which include the kinetic theory approach, quasiparticle model, Chapman-Enskog method, Gribov plasma, Polyakov-quark-meson model~\cite{Jaiswal:2020qmj, Sasaki:2008fg, Torrieri:2007fb, Plumari:2011mk, Soloveva:2019xph, Berrehrah:2016vzw, Mitra:2017sjo, Singha:2017jmq}. The transport properties of the QGP have also been extensively studied in the high-temperature regime using resummed perturbation theory~\cite{Andersen:2004fp} as well as lattice calculations~\cite{HotQCD:2014kol, Meyer:2007dy, Meyer:2007ic}. In the phenomenologically relevant temperature range, typically $T_c\lesssim T \lesssim 4T_c$, where $T_c$ represents the critical temperature, it is evident that the perturbative calculation alone is not an efficient tool due to the limitations that arise when the coupling strength of the system becomes of the order of unity~\cite{Su:2015esa}. Due to this, a theoretical model that encapsulates the non-perturbative features is more favoured for exploring the medium QGP.
	
Lattice QCD is a numerical approach that is best suited for reliable computations in the nonperturbative regime. However, lattice QCD is computationally expensive, and it is challenging to calculate dynamical quantities. Moreover, past studies based on lattice calculation have revealed that the results for transport coefficients are plagued by large inconsistencies, particularly in the phenomenologically relevant temperature regime~\cite{Meyer:2007dy, Meyer:2007ic} and are still inconclusive. Therefore, it is useful to have an alternative approach to incorporate nonperturbative features in the theory. In this work, we adopt the prescription suggested by Gribov in his seminal work Ref.~\cite{Gribov:1977wm}. Following this idea, Zwanziger, in the later year, formulated the local and renormalizable action discussed in detail in Ref.~\cite{Zwanziger:1989mf}. 
	
Gribov's prescription has been explored quite extensively encompassing a broad spectrum of topics in Refs.~\cite{Bandyopadhyay:2015wua, Begun:2016lgx, Su:2014rma, Dudal:2017kxb, Bandyopadhyay:2023yjp, Wu:2022nbv, Canfora:2015yia, Gotsman:2020ryd, Debnath:2023dhs, Debnath:2023zet, Sumit:2023hjj, Sumit:2023oib}. In~\cite{Kharzeev:2015xsa}, the connection between the Gribov parameter and the topological structure of the QCD vacuum has been explored. The results of this study shed light on a profound relationship between the Gribov-Zwanziger (GZ) quantization and the confinement-deconfinement phase transition. Earlier works, such as Refs.~\cite{Madni:2022bea, Fukushima:2013xsa, Dudal:2008sp, Burgio:2008jr}, have yielded promising agreement with lattice results, particularly in the infrared regime. The compelling agreement with lattice results has motivated us to incorporate Gribov-modified gluons in our investigation of the transport coefficients of QGP. To further explore the impact of quarks on these transport coefficients, we employed a quasiparticle approach~\cite{Gorenstein:1995vm} to formulate the thermodynamics of quarks, ensuring thermodynamic consistency throughout. Subsequently, using kinetic theory within the relaxation time approximation, we derived expressions for both shear and bulk viscosities of the QGP which consist of Gribov modified gluons and quasiparticle quarks. A similar approach has been explored previously in Ref.~\cite{Mykhaylova:2020pfk}, where the influence of dynamical quarks on QGP transport coefficients was explored by incorporating the interactions within the effective mass through a temperature-dependent running coupling derived from lattice QCD thermodynamics.

The structure of this article is organized as follows: In Sec.~\ref{sec:gribovgluons}, we introduce the Gribov quantization approach for non-Abelian gauge fields, providing a brief discussion on the motivation behind this idea. Section~\ref{sec:formalism} focuses on the thermodynamics of the gluonic plasma, employing the kinetic theory framework combined with the Gribov prescription. Simultaneously, we employ the quasiparticle approach to analyze the thermodynamic properties of quarks, ensuring consistency within the relevant temperature range. The interaction between quarks and gluons in the QGP is encapsulated in the quasi-mass of quarks, denoted as $m_q$. In Sec.~\ref{sec:vssq}, we examine the speed of sound square ($c_s^2$), which plays a crucial role in determining the bulk viscosity of the system. In Sec. \ref{subsec:transportcoe}, building on the theoretical framework established in Ref.~\cite{Chakraborty:2010fr}, we derive expressions for the shear and bulk viscosities, first for the quasi-quarks and subsequently for the Gribov-modified gluons. Our analysis demonstrates that the inclusion of quasi-quarks significantly suppresses the scaled transport coefficients across the entire temperature range under consideration ($1 \leq T/T_c \leq 3.5$). Finally, in Sec.~ \ref{sec:summary}, we provide a comprehensive summary of our findings and offer concluding remarks. We also discuss broader implications of our work and highlighting possible directions for future research. This structured approach allows us to systematically address the influence of non-perturbative effects on QGP transport properties, particularly near the critical temperature, where significant changes in the medium occur.
	
Throughout this paper, we adopt natural units with $c=\hbar=k_B=1$. We denote the magnitude of three vectors using bold and italic font (\textbf{\textit{p}}), while four vectors are represented in a standard font. The center-dot notation signifies the scalar product of four vectors employing the metric $g^{\mu\nu}={\rm diag}(+1,-1,-1,-1)$. The Minkowski four-vector $p^\mu=(p_0,\vec{\textbf{\textit{p}}})$, and the magnitude of $\vec{\textbf{\textit{p}}}$ is denoted by $\textbf{\textit{p}}$. Additionally, the Euclidean four-vector is denoted as $P^\mu=(p_4,\bm{p})$.
\section{Gribov Gluons}
\label{sec:gribovgluons}
The gauge fixing method in the quantization of the non-abelian gauge field was first proposed by Faddeev and Popov back in 1967~\cite{Faddeev:1967fc}, where the gauge fixing was ensured by incorporating the delta function into the path integral which, without the source term in Euclidean space becomes~\cite{Sainapha:2019cfn}
	\begin{align}
		\mathcal{Z}(0)=\int  \mathcal{D}\Lambda  \int \mathcal{D}A \hspace{0.15cm}\delta(\mathcal{G}(^{\Lambda}\!A))e^{-S_{YM}}\Delta_\text{FP}~.
		\label{Z0}
	\end{align}
	In Eq.~\eqref{Z0}, $S_{YM}$ is the Yang-Mills action, $\int \mathcal{D}A$ denotes the integral over the gauge field configurations, $\Delta_\text{FP}$ is the Faddeev-Popov determinant given as
	\begin{align}
		\Delta_{FP}=\bigg|\frac{1}{g} {\rm det}\mathcal{M}_{ab}\bigg| =\bigg|{\rm det}\frac{\delta \mathcal{G}^a(^{\Lambda}\!A))}{\delta \Lambda^b}\bigg|_{\mathcal{G}(^{\Lambda}\!A)=0}~,
		\label{deltafp}
	\end{align}
	where $\mathcal{G}(^{\Lambda}\!A)$ is the gauge constraint. Additionally, $\Lambda^b$ is the infinitesimal parameter of a particular gauge transformation $U=1-\Lambda^bT^b+\mathcal{O}(\Lambda^2)$. The Faddeev-Popov operator in Euclidean space is given as
\begin{align}
		\mathcal{M}^{ab}(x,y)=-\partial_{\mu} D_{\mu}^{ab}\delta(x-y)~.
		\label{mab}
\end{align}
In Eq.~\eqref{mab}, $D_{\mu}^{ab}$ is the covariant derivative. While the Faddeev-Popov quantization has shown considerable success across various aspects, it is still not free from the ambiguity as pointed by Gribov in Ref.~\cite{Gribov:1977wm}. The root of the issue lies in the ideal gauge fixing condition by specifying that the gauge orbit must intersect the gauge fixing constraint surface precisely once. However, in practical scenarios, this is not true, perticularly in the non-perturbative domain and is identified as Gribov ambiguity. Gribov's approach, extensively reviewed in Refs.~\cite{Vandersickel:2012tz, Sobreiro:2005ec}, focuses on improving the infrared (non-perturbative) dynamics of non-abelian gauge theory by fixing the residual gauge transformations which exists even after applying Faddeev-Popov quantization. Mathematically, it was achieved by restricting the path integral to a Gribov region $\Omega$ in a way so chosen that for any gauge field configuration $A_{\mu}^a $,
	\begin{align}
		\Omega=\left\{A^{a}_{\mu}, \hspace{0.1cm} \partial_{\mu}A_{\mu}^a = 0;\hspace{0.1cm} \mathcal{M}^{ab}>0 \right\}~,
	\end{align} 
where the gauge field $A_{\mu}^a$ obeys the Landau gauge condition for which the Faddeev-Popov operator $\mathcal{M}_{ab}$ is positive definite to ensure the no-pole condition required the overcome the Gribov ambiguity. Following this restriction on the gauge field configuration, the partition function in the Euclidean space-time reads as~\cite{Vandersickel:2012tz}
	\begin{align}
		\mathcal{Z}&=\int [\mathcal{D}c] \int [\mathcal{D} \bar{c}] \int [\mathcal{D}A] \hspace{0.1cm}V(\Omega)\delta(\partial\! \cdot\! A) \nonumber \\
		& \times {\rm exp}\hspace{0.1cm}[-S_{YM}-\int d^4x \bar{c}^a(x)\partial_{\mu}D_{\mu}^{ab}c^b(x)]~,
		\label{zgribov}
	\end{align}
where $c(x)$ and $\bar{c}(x)$ are the ghost and anti-ghost fields respectively. The term $V(\Omega)=\theta[1-\sigma_0]$ is the step function needed to take care of the no-pole condition. Also, $[1-\sigma_0]=[1-\sigma(P=0)]$ is the inverse of ghost dressing function, defined as $Z_G=[1-\sigma(P)]^{-1}$. To obtain the gluon two-point function, the constraint on the gauge fields is inserted in the Eq.~\eqref{zgribov} by re-expressing the parameter $V(\Omega)$ as 
	\begin{align}
		V(\Omega)=\frac{1}{2\pi i}\int_{-i \infty+0^+}^{i\infty+0^+}\frac{ds}{s}\hspace{0.1cm}{\rm \exp}[s(1-\sigma_0)]~.
	\end{align}
In the original Gribov quantization, the introduction of the Gribov parameter effectively modifies the gluon propagator, leading to resolution of the Gribov ambiguity by limiting the gauge field configurations. However, this approach has inherent limitations. In this context, the need for further refinement arises, motivating the transition to the Refined Gribov-Zwanziger (RGZ) framework. The RGZ approach incorporates additional degrees of freedom via auxiliary fields, leading to the inclusion of mass terms (corresponding to the existance of the non-zero value of the condensate $\langle A^2 \rangle$) in the propagator (see Refs.~\cite{Sobreiro:2004us, Dudal:2008sp} for further details). In the persence of the dynamical mass term $m_g$, the Gribov-modified gluon propagator in Landau gauge takes the form
	\begin{align}
		\mathfrak{D}^{ab}_{\mu\nu}(P)=\delta^{ab}\frac{P^2}{P^4+m_g^2P^2+\gamma_G^4}\left[\delta_{\mu\nu}-\frac{P_{\mu}P_{\nu}}{P^2}\right]~,
		\label{propagator}
	\end{align} 
where  $\gamma_G$ represents the Gribov parameter. Interestingly, $m_g$ and $\gamma_G$ are not independent but are linked through the gap equation, which at one-loop order takes the form Ref.~\cite{Sobreiro:2004us} as 
	\be
	\sumint_{P} \frac{1}{P^4+m_g^2 P^2+\gamma_G^4}=\frac{d}{(d-1)N_cg^2},
	\label{gap_eq}
	\ee
where $N_c$ is the number of colors (here 3) and $g$ is the running coupling, $d$ is for the dimension of the space-time. It is also worth to realize that at one-loop, the analytical solution at asymptotically high temperature $T$ gives us~\cite{Zwanziger:2006sc}
	\begin{align}
		\gamma_G\sim g^2 T,
		\label{gribovparameter}
	\end{align} 
It is evident from Eq.~\eqref{gribovparameter} that the Gribov parameter is proportional to the magnetic scale, i.e. $\gamma_G \propto g^2T$. Interestingly, the inclusion of the magnetic scale inherently suggests a promising resolution to IR catastrophe caused by the magnetic scale, called the Linde problem (see~\cite{Linde:1980ts} for detail). Therefore, the dynamical gluonic mass along with the Gribov parameter of magnetic scale ($g^2T$) could be utilized to explore the dynamics of the deconfined medium along with the temperature dependence of its transport coefficients. A similar approach has been previously studied in Ref.~\cite{Bandyopadhyay:2015wua} where dilepton rate and quark number susceptibility have been worked out and also in Ref.~\cite{Su:2014rma} to examine the quark self-energy at one-loop level at high temperature. 
\section{Formalism}\label{sec:formalism}
\subsection{Thermodynamics}\label{subsec:thermodynamics}
In our investigation of the thermodynamics of the medium, we have employed the well-established kinetic theory approach~\cite{Jeon:1994if}. We have dedicated separate sections to develop the thermodynamics of quasiparticle quarks and Gribov-modified gluons, adopting distinct methodologies for each. 
	
To characterize the system's dynamics, the energy-momentum tensor takes on the following form~\cite{Jeon:1995zm}
	\begin{align}
		T_{(0)}^{\mu\nu}= \int dp \cdot p^{\mu}p^{\nu} f_0+B_0(T)g^{\mu\nu}~,
		\label{energymomentumtensor}
	\end{align}
where $B_0(T)$ is bag pressure, which is added to take care of thermodynamic consistency in equilibrium and $f_0$ is the local equilibrium distribution function. 

The Lorentz invariant momentum integral in the Minkowski space-time for Gribov gluons/quasiparticle(q) quarks with quasi mass $m_q$ is
\begin{align}
	\int dp_{gl/q}=\frac{d_{gl/q}}{(2\pi)^3}\int d^3\textbf{\textit{p}} \int dp_0 \hspace{0.1cm}  2 \Theta(p_0) \hspace{0.1cm} \delta(\mathcal{K}_{gl/q})~,
	\label{dpgluon}
\end{align}
where $\mathcal{K}_{gl}=(p^2-m_g^2+\gamma_G^4/p^2)$ and $\Theta(p_0)$ is the Heaviside step function to ensure that only the states with positive energy are considered. Also, for the case of the quasiparticle quarks, $\mathcal{K}_{q}=(p^2-m_q^2)$. The degeneracy factor is given by $d_{gl/q}$.

The pressure and energy density are expressed as~\cite{Jaiswal:2020qmj}
	\begin{eqnarray} 
		\mathcal{P} &=& -\frac{1}{3}\Delta_{\mu\nu}T_{(0)}^{\mu\nu}=\mathcal{P}^{Kinetic}-B_0(T)~
\label{pressure}\\
		\mathcal{E} &=& u_{\mu}u_{\nu}T_{(0)}^{\mu\nu}=\mathcal{E}^{kinetic}+B_0(T)~.
		 \label{energy}
	\end{eqnarray}
where $\Delta^{\mu\nu}=g^{\mu\nu}-u^{\mu}u^{\nu}$ is the projection tensor normal to fluid four velocity $u^{\mu}$, satisfying $u^{\mu}u_{\mu}=1$. In fluid rest frame, $u^{\mu}=(1,0,0,0)=(1,\vec{0})$. $\mathcal{P}^{Kinetic}$ and $\mathcal{E}^{kinetic}$ are the particle contribution to pressure and energy density, respectively.	
	\subsubsection{Gribov gluons}
	\label{subsec:gribovgluons}
The thermodynamics (also the transport coefficients) of gluonic plasma is equipped with two unknown parameters, the Gribov parameter and the dynamical mass. To accurately determine their temperature dependence, two independent equations are required. In this study, we first solved the gap equation (Eq.~\ref{gap_eq}) numerically, obtaining $m_g$ as a function of $\gamma_G$, which is to say $m_g\sim m_g(\gamma_G)$. To further constrain the temperature dependence of these variables, we utilized the lattice equation of state to derive the temperature dependence of $\gamma_G$, and by extension, the dynamical mass $m_g$. For the finite temperature, the integral in the Eq.~\eqref{gap_eq} has been solved within the ($\overline{\text{MS}}$) renormalization scheme with $d=4-2\varepsilon$ and 
	\begin{align}
		\sumint_{P}=\left(\frac{e^{\gamma_E}\mu^2}{4\pi}\right)^{\varepsilon} \cdot \left( T \sum_{p_4=2\pi n T}\right)\cdot \int \frac{d^{3-2\epsilon}\textbf{\textit{p}}}{(2\pi)^{3-2\varepsilon}}~.
		\label{msbarsumint}
	\end{align} 
The expression within the bracket represents the Matsubara frequency summation, where the index $n$ takes on all integer values, ranging from $-\infty$ to $\infty$, in discrete steps of one. This summation is a critical component in finite-temperature quantum field theory, capturing the thermal effects by summing over the Matsubara frequencies, which correspond to the allowed discrete energy levels of the system. All calculations were performed with the renormalization scale $\mu$ specifically chosen as $\mu = 4\pi T$. Through this framework, the temperature-dependent behavior of the Gribov parameter and dynamical mass is determined.
	
Solving the gap equation [Eq.~\eqref{gap_eq}] using the Eq.~\eqref{msbarsumint} in $\overline{\text{MS}}$ scheme, we get
	\begin{align}
		\frac{3 N_c g^2}{128 \pi^2\sqrt{m_g^4-4\gamma_G^4}}\left[\chi_1+\chi_2+\chi_3\right]=1~,
		\label{finalgapequation}
	\end{align}
	where
	\begin{align}
		\chi_1=\left(\sqrt{m_g^4-4\gamma_G^4}-m_g^2\right)\left(\frac{5}{6}-\ln \frac{m_g^2-\sqrt{m_g^4-4\gamma_G^4}}{2\mu^2}\right),\!
		\label{chi1}
	\end{align}
	\begin{align}
		\chi_2=\left(\sqrt{m_g^4-4\gamma_G^4}+m_g^2\right)\left(\frac{5}{6}-\ln \frac{m_g^2+\sqrt{m_g^4-4\gamma_G^4}}{2\mu^2}\right),\!
		\label{chi2}
	\end{align}
and
		\begin{align}
		\chi_3=16 \int_0^{\infty} d\textbf{\textit{p}}{\hspace{0.075cm}}\textbf{\textit{p}}^2 {\hspace{0.075cm}} \left(\frac{f_0^{gl}(\epsilon_1)}{\epsilon_1}-\frac{f_0^{gl}(\epsilon_2)}{\epsilon_2}\right).
		\label{chi3}
	\end{align}
In the absence of the dynamical mass term $(m_g \rightarrow 0)$, the above form of the gap equation simplifies to the same form as derived in Ref.~\cite{Fukushima:2013xsa}, where they solved it without considering this term. In Eq.~\eqref{chi3}, $f_0^{gl}$ is the Bose equilibrium distribution function given as
	\begin{align}
		f_0^{gl}(\epsilon)=\left[{\rm exp}\left(\frac{\epsilon}{T} \right) -1 \right]^{-1}~.
		\label{distributionfngluon}
	\end{align}	
Additionally, and $\epsilon_1$ and $\epsilon_2$ are the energy corresponding to the pole of the Gribov-modified gluon propagator defined in Eq.~\eqref{propagator} whose forms are 
		\begin{align}\label{epsilon1}
			\epsilon_1=\sqrt{\textbf{\textit{p}}^2+m_1^2}\\
	\label{epsilon2}
	\epsilon_2=\sqrt{\textbf{\textit{p}}^2+m_2^2},
		\end{align}		
	with 
	\begin{eqnarray}
		m_1^2=\frac{1}{2}\left(m_g^2-\sqrt{m_g^4-4\gamma_G^4}\right)\\
			m_2^2=\frac{1}{2}\left(m_g^2+\sqrt{m_g^4-4\gamma_G^4}\right).
		\end{eqnarray}
%
In Eq.~\eqref{finalgapequation}, $g$ is the running coupling we have taken it from Ref.~\cite{Fukushima:2013xsa} as
	\begin{align}
		\frac{g^2(T/T_c)}{4\pi}=\frac{6 \pi}{11 N_c \ln[c_{IR}(T/T_c)]}~,
	\end{align}
with $T_c=260$ GeV being the critical temperature for the gluonic plasma and $c_{IR}=1.43$ is obtained by extracting the running coupling from the large-distance (IR) behaviors of the heavy-quark free energy~\cite{Fukushima:2013xsa}. Now, to get the temperature dependence of the Gribov parameter, we formulate the thermodynamics of the gluonic plasma based on the framework described in section~\ref{sec:gribovgluons}.
	
Using Eqs.~\eqref{pressure},~\eqref{energy} and~\eqref{dpgluon}, for gluonic plasma, the pressure and energy density are obtained as
	\begin{equation} 
		\begin{split}
			\mathcal{P}_{gl}&= -\frac{1}{3}\Delta_{\mu\nu}T_{(0)}^{\mu\nu}=\mathcal{P}_{gl}^{Kinetic}-B_0^{gl}(T) \\
			& =\frac{d_{gl}}{3\pi^2} \int\! d\textbf{\textit{p}} \hspace{0.05cm} \textbf{\textit{p}}^4 \hspace{0.05cm} \left[\frac{f^0_{gl}(\epsilon_1)}{\mid f^\prime (\epsilon_1) \mid}+\frac{f^0_{gl}(\epsilon_2)}{\mid f^\prime (\epsilon_2) \mid}\right]-B_0^{gl}(T) ~.
		\end{split}
	\label{pressuregluon}
	\end{equation}
	\begin{eqnarray} 
			\mathcal{E}_{gl} &= &u_{\mu}u_{\nu}T_{(0)}^{\mu\nu}=\mathcal{E}_{gl}^{kinetic}+B_0^{gl}(T)\nonumber
			\\
			&&\hspace{-0.8cm}=\frac{d_{gl}}{\pi^2}\int d\textbf{\textit{p}} \hspace{0.05cm} \textbf{\textit{p}}^2 \left[\frac{\epsilon_1^2 f^0_{gl}(\epsilon_1)}{\mid f^\prime(\epsilon_1) \mid}+\frac{\epsilon_2^2 f^0_{gl}(\epsilon_2)}{\mid f^\prime(\epsilon_2) \mid} \right]+B_0^{gl}(T) . \! 
	\label{energygluon}
	\end{eqnarray}
$d_{gl}$ is the degeneracy factor for gluonic plasma, which is $d_{gl}=2\times(N_c^2-1)=16$. 

Note that the bag pressure $B_0^{gl}(T)$ is added to the energy density and subtracted from the pressure to maintain the thermodynamics consistency. Also, in Eqs.~\eqref{pressuregluon} and~\eqref{energygluon}, we have
		\begin{align}
			f^\prime(\epsilon_1)&=\frac{4\epsilon_1\left(m_g^4-4\gamma_G^4-m_g^2\sqrt{m_g^4-4\gamma_G^4}\right)}{\left(m_g^2-\sqrt{m_g^4-4 \gamma_G^4}\right)^2}~\nonumber\\
			&=-\frac{\epsilon_1}{\gamma_G^4}\left(m_g^4-4\gamma_G^4+m_g^2\sqrt{m_g^4-4\gamma_G^4}\right),
		\end{align}
		and
		\begin{align}
			f^\prime(\epsilon_2)&=\frac{4\epsilon_2\left(m_g^4-4\gamma_G^4+m_g^2\sqrt{m_g^4-4\gamma_G^4}\right)}{\left(m_g^2-\sqrt{m_g^4-4 \gamma_G^4}\right)^2}~\nonumber\\
			&=-\frac{\epsilon_2}{\gamma_G^4}\left(m_g^4-4\gamma_G^4-m_g^2\sqrt{m_g^4-4\gamma_G^4}\right),
		\end{align}
Given the expressions for the pressure in Eq.~\eqref{pressuregluon} and the energy density in Eq.~\eqref{energygluon}, the entropy density can be determined using the thermodynamic relation $s_{gl} = (\mathcal{P}_{gl} + \mathcal{E}_{gl})/T$. It is important to note that this expression for entropy density is independent of any bag correction term. Since the derived form of the entropy density explicitly involves the Gribov parameter as a variable, we proceed by employing lattice data for the equation of state of the gluonic plasma to determine the temperature dependence of the Gribov parameter and, subsequently, the dynamical mass \( m_g \equiv m_g(\gamma_G) \), as obtained by solving the gap equation.

For the gluonic plasma, we utilize the lattice data of the equation of state (pressure, trace anomaly, energy density, enropy density) available in the Ref.~\cite{Borsanyi:2012ve}. In order to fit the scaled trace anomaly $(\mathcal{I}/T^4)$ with the lattice findings, we have made use of the following fit function, also explored in Ref.~\cite{Borsanyi:2010cj}
	\begin{align}
		\frac{\mathcal{I}}{T^4} = & \exp\left[-\left(\frac{h_1}{\hat{T}}+\frac{h_2}{\hat{T}^2}\right)\right] \nonumber \\
		&\times\left(\frac{h_0}{1+h_3 \hat{T}^2}+\frac{f_1({\rm \tanh}[f_2\hat{T}+f_3]+1)}{1+g_1\hat{T}+g_2\hat{T}^2} \right),
		\label{TAPG}
	\end{align} where $\hat{T}$ is the scaled temperature ($=T/T_c$). 
	
\begin{figure}
		\centering
		\includegraphics[scale=0.49]{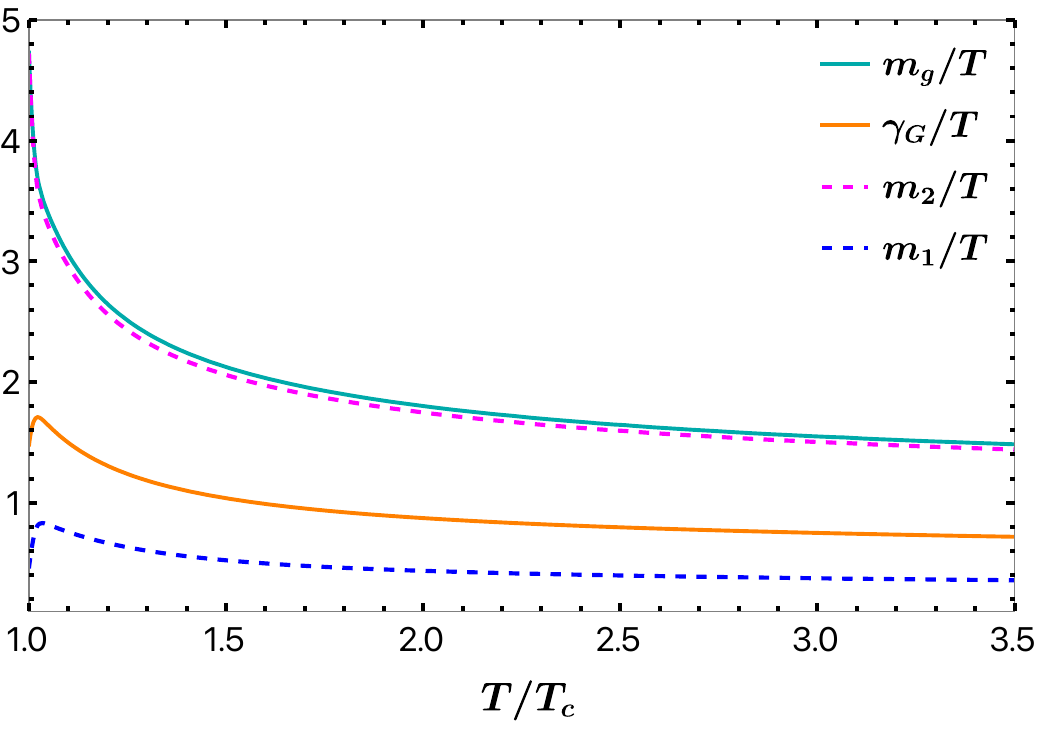}
		\caption[Fitted entropy]{The scaled dynamical mass ($m_g/T$) and the scaled Gribov parameter ($\gamma_G/T$), the scaled gluon quasiparticle masses $(m_{1,2}/T)$ are plotted as a function of scaled temperature $(T/T_c)$ in the temperature range ($ 1 \le T/T_c \le 3.5 $).}
		\label{fig:mgandgammag}
	\end{figure}

For the set of fitted parameters from Table~\ref{table}, the scaled pressure for $N_f=0$ can be obtained by the expression
	\begin{align}
		\frac{\mathcal{P}(T)}{T^4}=\frac{\mathcal{P}(T_0)}{T^4}+\int_{T_0}^{T}\frac{dT}{T}\times \frac{\mathcal{I}}{T^4} ~,
		\label{fittedpressuregluon}
	\end{align} 
where ($\mathcal{P}(T_0=0.7 T_c)/T^4=0.0015$). The energy density is a crucial parameter for understanding the complete thermodynamic behavior of the medium under consideration. From Eq.~\eqref{TAPG} and~\eqref{fittedpressuregluon}, the relation ($\mathcal{I} = \mathcal{E} - 3\mathcal{P}$) provides the variation of the energy density ($\mathcal{E}$) as a function of the scaled temperature $T/T_c$. To obtain the entropy density, we use the thermodynamic relation ($s = d\mathcal{P}/dT =(\mathcal{P}+\mathcal{E})/T)$. Finally, the entropy density of the Gribov-modified gluon is matched to the lattice entropy density to fit the Gribov parameter as a function of the scaled temperature, shown in Fig.~(\ref{fig:mgandgammag}). 
It is essential to note that the bag correction term for the gluonic plasma has been determined by enforcing thermodynamic consistency through the relation $s=d\mathcal{P}/dT =(\mathcal{P}+\mathcal{E})/T$, which matches with the fundamental thermodynamic identity. This condition ensures that the pressure (Eq.~\ref{energygluon}) and energy density (Eq.~\ref{pressuregluon}) evolve in a physically consistent manner with respect to temperature. 
\subsubsection{Quasi quarks}
The quasiparticle approach is an effective description and provides a useful way to understand and model the complex dynamics of strongly interacting systems under extreme conditions. In the quasi-particle approach, the energy of the particle (here, (anti)quarks) is taken as  $\epsilon_q=(\textbf{\textit{p}}^2+m_q^2)^{1/2}$, where $m_q\equiv m_q(T)$ denotes the temperature-dependent mass of the quasi quark. The local equilibrium distribution function of the quark, following the Fermi Dirac statistics, is given as
	\begin{align}
		f_0^q(\epsilon_q)=\left[{\rm exp}\left(\frac{\epsilon_q}{T} \right) +1 \right]^{-1}~.
		\label{fermidistribution}
	\end{align}
\begin{figure}
	\centering
	\includegraphics[scale=0.66]{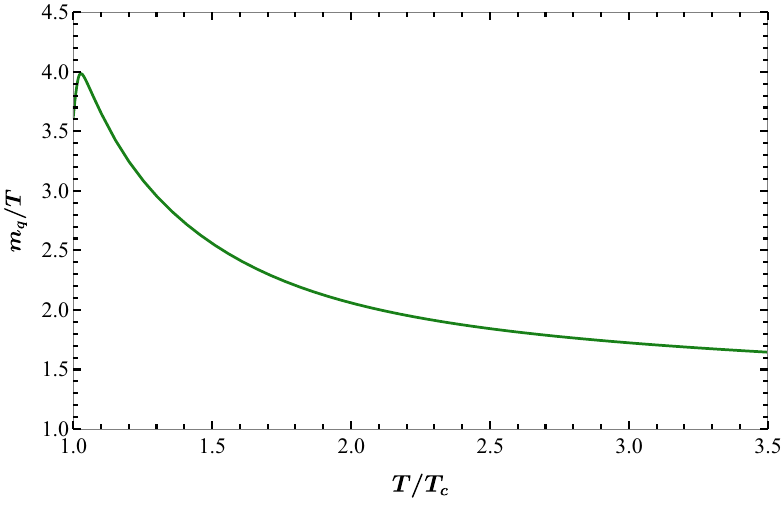}
	\caption{Quasi-quark mass, $m_q$, scaled with temperature $T$, plotted as a function of scaled temperature in the range ($1.0 \le T/T_c \le 3.5$).}
	\label{fig:mq}
\end{figure}
In order to formulate the thermodynamics of quasi-quarks, the pressure and energy density are evaluated using Eq.~\eqref{pressure} and Eq.~\eqref{energy}.

Using the similar methods discussed in the previous subsection, we obtain respectively the  pressure ($\mathcal{P}_q$) and energy density ($\mathcal{E}_q$) for the quasi-quarks as
	\begin{align}
		\mathcal{P}_q=\frac{d_{q}}{6\pi^2} \int d\textbf{\textit{p}} \hspace{0.05cm}\textbf{\textit{p}}^4  \frac{f^0_q(\epsilon_q)}{\epsilon_q} -B_0^{q}(T)~,
		\label{pressurequark}\\
		\mathcal{E}_q=\frac{d_{q}}{2\pi^2} \int d\textbf{\textit{p}} \hspace{0.05cm}\textbf{\textit{p}}^2 f^0_q(\epsilon_q)\epsilon_q+B_0^q(T)~.
		\label{energyquark}
	\end{align}
The degeneracy factor, in this case, is $d_{q}=2 \times 2\times { N_c\times N_f}$, where the factors correspond to the contributions from anti-quarks, spin, number of colors, and flavors, respectively. Also, $B_0^q(T)$ is the bag-like term in quasi-quarks pressure and energy density. In order to ensure thermodynamic consistency in the calculation, it is required that $s=d\mathcal{P}_q/dT=(\mathcal{P}_q+\mathcal{E}_q)/T$. Therefore, taking the derivative of the pressure from Eq.~\eqref{pressurequark}, it is obtained that the thermodynamic consistency is maintained when	
	\begin{align}\label{thermoconsistency}
		\frac{dB_0^q(T)}{dT}+\frac{d_{q}}{2\pi^2} m_q m_q^\prime  \displaystyle\int d\textbf{\textit{p}} \hspace{0.12cm} \textbf{\textit{p}}^2  \frac{f^0_q(\epsilon_q)}{\epsilon_q} =0~,	
	\end{align}
where $m_q^\prime=dm_q/dT$. 

It should be noted here that in the present framework, the quark-gluon interaction is captured in the quasi-mass $m_q$. In order to obtain the temperature dependence of the quasi mass $m_q$, we equate the entropy density of the lattice finding of (2+1)QCD to that of the calculated entropy density by summing up the pressure and energy density using the Eqs.~\eqref{pressuregluon},~\eqref{energygluon},~\eqref{pressurequark} and~\eqref{energyquark}, i.e.
	\begin{align}
		\sum_{g,q,\bar{q}} s= s_{(2+1) QCD}^{ Lattice}~,
		\label{stotal}
	\end{align}
where $g$, $q$ and $\bar{q}$ refers to the gluons, quarks and anti-quarks respectively and $T$ is the entropy density in the quasi-particle approach. The lattice data considering the quark effect is taken from Ref.~\cite{Borsanyi:2010cj}.	

\begin{figure}
	\centering
	\includegraphics[scale=0.535]{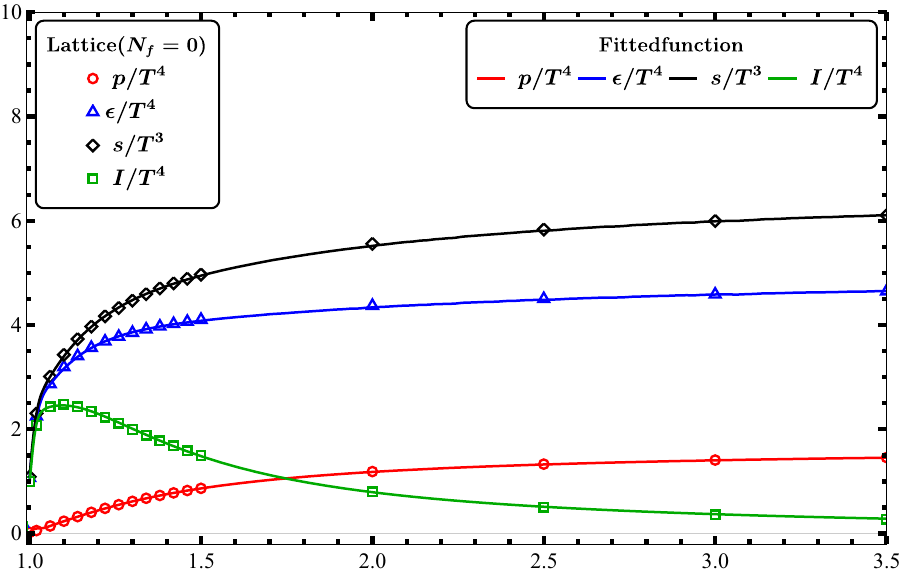}
	\includegraphics[scale=0.535]{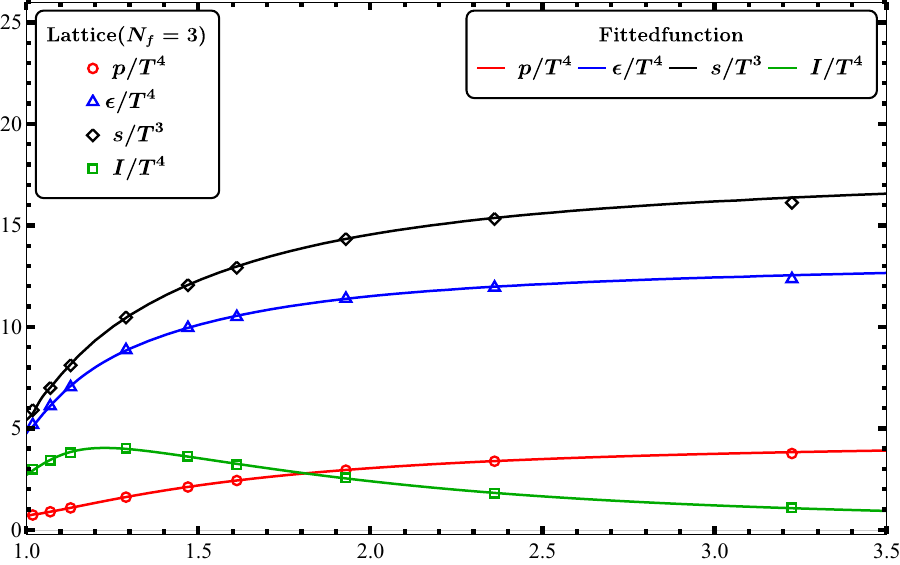}
	\caption{Thermodynamical variables (free energy, the energy density and the entropy density) from the Eqs.~\eqref{TAPG} and~\eqref{fittedpressuregluon} with fitted parameters given in Table~\ref{table} for $N_f=0$ as well as $N_f=3$ along with corresponding lattice findings of the same.  $T_c$ for $N_f=0$ is taken as $260$ MeV while for $N_f=3$ is $155$ MeV.}
	\label{fig:lattice}
	\end{figure}

In this case as well, the functional form of the fit function of the trace anomaly and the scaled pressure is same as give in the Eqs.~\eqref{TAPG} and~\eqref{fittedpressuregluon}, respectively, with different set of fit parameters shown in the Table~\ref{table}. Equipped with the pressure and trace anomaly fit functions corresponding to lattice data available in Ref.~\cite{Borsanyi:2010cj}, the sum of entropy densities of quark (with quasi mass $m_q$) and gluons has been equated with that of the entropy density of the medium QGP, like Eq.~\eqref{stotal}. As a result, we got the temperature dependence of scaled quasi mass $m_q(T)/T$, shown in Fig.~(\ref{fig:mq}).
	\begin{table}[ht]
		\centering
		\renewcommand{\arraystretch}{1.5} 
		\setlength{\tabcolsep}{25pt} 
		\begin{tabular}{|c|c|c|}
			\hline
			\multicolumn{3}{|c|}{\textbf{Fit parameters}} \\
			 \hline
			 & $N_f=0$ & $N_f=3$ \\
			\hline
			$h_0$ & 0.23 & 0.28 \\
			\hline
			$h_1$ & -1.83 & 0.41 \\
			\hline
			$h_2$ & 2.92 & -0.34 \\
			\hline
			$h_3$ & 0.07 & 0.002 \\
			\hline
			$f_1$ & 0.32 & 1.50 \\
			\hline
			$f_2$ & 62.39 & 6.04 \\
			\hline
			$f_3$ & -62.55 & -5.96 \\
			\hline
			$g_1$ & -1.98 & -0.87 \\
			\hline
			$g_2$ & 1.08 & 0.49 \\
			\hline 
\end{tabular}
\caption{The values of the fit parameters in the Eq.~\eqref{TAPG} as a result of fitting with the lattice data of the scaled trace anomaly ($\mathcal{I}/T^4$) as a function of scaled temperature ($T/T_c$) available in Ref.~\cite{Borsanyi:2012ve}. While for $N_f=3$, parameter values were taken from Ref.~\cite{Borsanyi:2010cj}.}
		\label{table}
	\end{table}
With the obtained values of the fit parameters in the Table~\ref{table}, the thermodynamic variables are plotted as a function of the scaled temperature (solid line) in the Fig.~(\ref{fig:lattice}) for the gluonic plasma as well as for the medium QGP along with the corresponding lattice findings. A decent match with the lattice data can be observed across the relevant temperature range.

Note that in the investigation of transport coefficients incorporating quasi-quark, we have made the significant assumption of treating light and strange quarks equally, despite their distinct physical characteristics, and neglected any possible dependence of transport coefficients on quark flavors.		
	\section{Speed of sound}
	\label{sec:vssq}
The square of speed of sound, $c_s^2$, plays an important role in characterizing the bulk viscosity and quantifying deviations from the conformal limit. Mathematically, $c_s^2$ is expressed as
	\begin{align}\label{cssq}
		c_s^2=\left(\frac{\partial \mathcal{P}}{\partial \mathcal{E}} \right)=\frac{s}{T}\left(\frac{\partial s}{\partial T}\right)^{-1}~.  
	\end{align}
In this study, we investigated two scenarios, one involving gluonic plasma ($N_f=0$) and the other being the medium QGP ($N_f=3$), consisting of quasi-quarks along with Gribov-modified gluons. The speed of sound square is explored in these two distinct medium and has been discussed in brief here. 
	
\begin{figure}
\centering
		\includegraphics[scale=0.48]{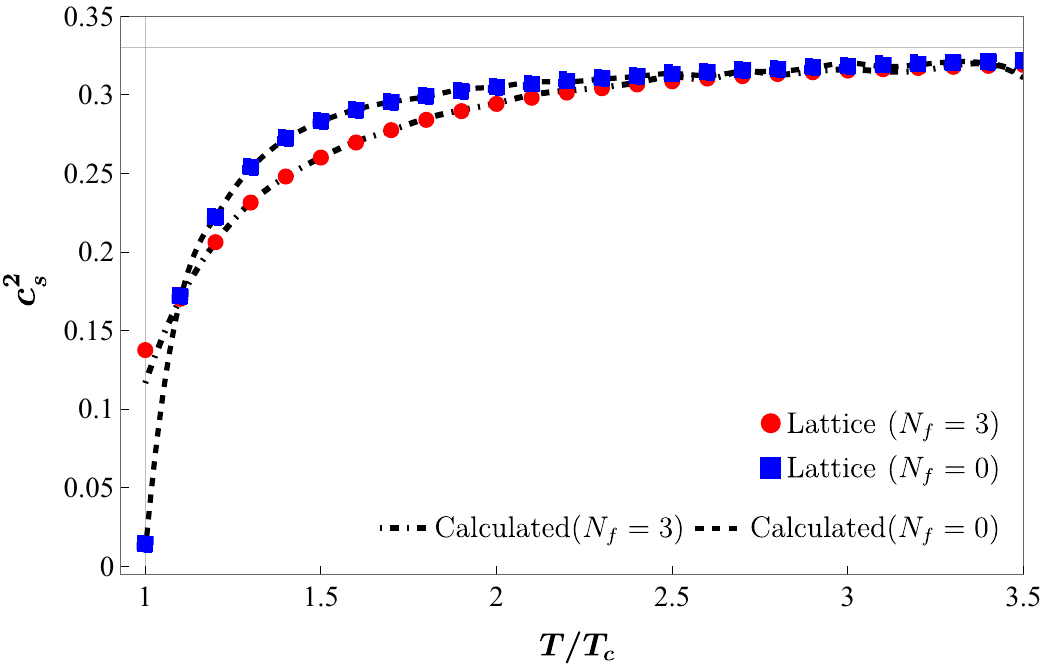}
		\caption[cssq]{Temperature dependence of the square of speed of sound, $c_s^2$ of Gribov-modified gluonic plasma as well as for QGP as a function of scaled temperature $T/T_c$. Lattice result for gluonic plasma has been obtained from Ref.~\cite{Borsanyi:2012ve}, while for $N_f=3$ has been taken from Ref.~\cite{Borsanyi:2010cj}. The horizontal solid grey line corresponds to the speed of sound $(c_s^2=1/3)$, which is attained in the conformal limit of the medium QGP.}
		\label{fig:cssq}
	\end{figure}

For the gluonic medium, the square of speed of sound is determined using Eqs.~\eqref{cssq},~\eqref{pressuregluon} and~\eqref{energygluon}. The temperature dependence of the same is depicted in Fig.~(\ref{fig:cssq}). Through our analysis, it is evident that the squared speed of sound in the gluonic plasma exhibits remarkable consistency above the critical temperature $T_c$ when compared with the one obtained from lattice data of the same. Also, we witness a pronounced escalation in $c_s^2$ near transition point, i.e., $T/T_c=1$. This notable increase can be attributed to a sudden discontinuity in the entropy density, indicative of a first-order phase transition in gluonic plasma.
	
Secondly, while adding the quasi-quark contribution, we have $c_s^2$ of the medium QGP. In this case, in order to evaluate the speed of sound square, we took the pressure ($\mathcal{P}$) and energy density ($\mathcal{E}$) as
\begin{eqnarray}
		\mathcal{P}&=&\mathcal{P}_{gl}+\mathcal{P}_{q}~,\\
			\label{pqcd}
		\mathcal{E}&=&\mathcal{E}_{gl}+\mathcal{E}_{q}~.
		\label{eqcd}
	\end{eqnarray}
Here we have used Eqs.~\eqref{pressuregluon},~\eqref{pressurequark},~\eqref{energygluon} and~\eqref{energyquark}. The speed of sound square of medium QGP is depicted in Fig.~(\ref{fig:cssq}) again, along with the lattice finding of the same. It is important to recognize that Eq.~\eqref{pqcd} represents an assumption within the present research model, where the total pressure, $\mathcal{P} = \mathcal{P}_{QCD}$, corresponds to the free energy density of the medium. This pressure is treated as the sum of contributions from the gluonic sector and the quasiparticle quark sector. In practice, however, at high temperatures, the free energy density of QCD exhibits cross-coupling effects regulated by the coupling $g$, and has been perturbatively obtained upto the order $\mathcal{O}(g^6 \ln(1/g))$ in Ref.~\cite{Kajantie:2002wa}.
It is clear from Fig.~(\ref{fig:cssq}) that our calculation matches the lattice finding in the temperature range $1 \le T/T_c \le 3.5$. Compared with the $c_s^2$ of purely gluonic medium (in the same figure), we notice in the present case that the variation with scaled temperature is smoother near the transition region. Also, we see that as the temperature increases, the speed of sound approaches the conformal limit, i.e., $c_s^2 \rightarrow 1/3$ for $N_f=0$ as well as for $N_f=3$. Also, apart from being an important input parameter in the transport coefficient(bulk viscosity), a decent match with the available lattice data of the speed of sound square to that of the calculated results solidifies the idea of thermodynamics being well settled for $N_f=0$ as well as for $N_f=3$. The next is to formulate the transport coefficients for the two cases.
	
	\section{Transport coefficients}\label{subsec:transportcoe}
In this section, we formulated the transport coefficients, specifically shear and bulk viscosities, for medium QGP following the theoretical framework developed in Ref.~\cite{Chakraborty:2010fr}. It is important to emphasize that this formulation is conducted in the limit of zero chemical potential. Furthermore, the analysis assumes the system to be in a state of near-local equilibrium, characterized by a spatially varying temperature distribution, $T(x)$, and a local flow velocity, $u(x)$.
	
For the medium slightly out of equilibrium, the distribution function $f(x,p)$ could be written as
\begin{align}
		f(x,p)=f^0(u\! \cdot\! p/T)[1+\phi(x,p)]~,
		\label{f}
\end{align}
It could also be realized that the deviation from the local equilibrium is related to the function $\phi(x,p)$ as
\begin{align}
		\delta f= f^{0}(u\! \cdot\! p/T)\phi(x,p)~.
		\label{deltaF}
\end{align}
Using Eqs.~\eqref{energymomentumtensor} and~\eqref{f}, the dissipative part of the energy-momentum tensor takes the form
	\begin{align}
		\Delta T^{\mu\nu}=\int \frac{d^3\textbf{\textit{p}}}{(2\pi)^3} \frac{p^{\mu}p^{\nu}}{\epsilon}f^{0}(u\! \cdot\! p/T)\times \phi(x,p),
		\label{deltaT1}
	\end{align}
where $|\phi(x,p)| \ll 1$. Additional constrain on $\phi(x,p)$ is realized by the Landau frame condition $u_{\mu}\Delta T^{\mu\nu}=0$. Also, the most general tensor structure of $\Delta T^{\mu\nu}$ could be expressed as, Ref.~\cite{Chakraborty:2010fr}
	\begin{align}
		\Delta T^{\mu\nu}=\eta\left(\nabla^{\mu}u^{\nu}+\nabla^{\nu}u^{\mu}-\frac{2}{3}\Delta^{\mu\nu}\partial_{\rho}u^{\rho}\right)+\zeta \Delta^{\mu\nu} \partial_{\rho}u^{\rho},
		\label{deltaT2}
	\end{align}
where $\nabla_{\alpha}= \Delta^{\alpha\beta}\partial_\beta= \partial_{\alpha}-u_{\alpha}(u \cdot \partial)$ is the derivative normal to the fluid velocity $u_{\alpha}$. Note that the terms $\eta$ and $\zeta$ in Eq.~\eqref{deltaT2} are the shear and bulk viscosities. In the similar fashion, the tensorial decomposition of the $\phi(x,p)$ is achieved as 
	\begin{align}
		\phi(x,p)=-\mathcal{A} \partial_{\rho}u^{\rho}+\mathcal{C}_{\mu\nu}\left(\nabla^{\mu}u^{\nu}+\nabla^{\nu}u^{\mu}-\frac{2}{3}\Delta^{\mu\nu}\partial_{\rho}u^{\rho}\right)\!. \hspace{-0.25cm}
		\label{phi}
	\end{align}
Furthermore, $\mathcal{C}_{\mu\nu}$ can be further decomposed as $\mathcal{C}_{\mu\nu}=\mathcal{C}p_{\mu}p_{\nu}$ because the contribution from the term $g_{\mu\nu}$ vanishes [see Ref.~\cite{Chakraborty:2010fr} for detail]. Substituting the Eq.~\eqref{phi} into the Eq.~\eqref{deltaT1} and then equating the dissipative part of the energy-momentum tensor from Eq.~\eqref{deltaT2}, we clearly see that in the local rest frame of the fluid, the bulk viscosity take the form
	\begin{align}
		\zeta=\frac{1}{3}\int \frac{d^3 \textbf{\textit{p}}}{(2\pi)^3}\frac{\textbf{\textit{p}}^2}{\epsilon}f^{0}(\epsilon)\mathcal{A}(\epsilon)~,
		\label{zeta}
	\end{align}
while the shear viscosity  
	\begin{align}
		\eta=\frac{2}{15}\int \frac{d^3\textbf{\textit{p}}}{(2\pi)^3}\frac{\textbf{\textit{p}}^4}{\epsilon}f^{0}(\epsilon)\mathcal{C}(\epsilon)~.
		\label{eta}
	\end{align}
It is evident from Eq.~\eqref{zeta} and Eq.~\eqref{eta} that to determine the temperature dependence of shear and bulk viscosities, it is essential to calculate the scalar coefficients $\mathcal{A}$ and $\mathcal{C}$. These coefficients can be obtained by solving the Boltzmann equation. In this work, we solved the Boltzmann equation by the relaxation time approximation (RTA) approach in order to determine the shear and bulk viscosities.
	
In the fluid rest frame, the Boltzmann equation in RTA is given as
	\begin{align}
		p^{\mu}\partial_{\mu}f(x,p)=-\frac{\epsilon\delta f(x,p)}{\tau_R}=-\frac{\epsilon f^0(x,p)\phi(x,p)}{\tau_R},
		\label{rtaBE}
	\end{align}
where $\delta f(x,p)$ is the deviation from the local equilibrium, defined in the Eq.~\eqref{deltaF}, $\tau_R$ is the relaxation time, which in this work has been assumed to be constant and same for the shear and bulk viscosity. The main task is now lies in the evaluation of the LHS of Eq.~\eqref{rtaBE}. In the view of the Eq.~\eqref{deltaF}, we then simplify the equation such that it matches with the Eq.~\eqref{phi}, which in turn helps us identify the coefficients $\mathcal{A}$ and $\mathcal{C}$. Both of these coefficients will have different forms for gluons and the quasiparticle quarks, which are discussed in the following separate subsections.
\subsection{Shear and bulk viscosity for quasiparticle quarks}
In this section, we formulate the expressions of shear and bulk viscosities for quasiparticle quarks. This is again, done by solving the Boltzmann equation in relaxation time approximation. Since the quasi mass $m_q$ is a function of temperature $T$, which in itself is a function of space-time ($T = T(x)$), therefore considering the mean field effects for the transport coefficients is required. We focus on the LHS of the Eq.~\eqref{rtaBE} and then with the help of Eq.~\eqref{phi}, we identify the coefficients $\mathcal{A}$ and $\mathcal{C}$. In the limit of small gradients (based on the assumption that $|\phi(x,p)|\ll 1$), solving the LHS of Eq.~\eqref{rtaBE}, we get
\begin{align}
		p^{\mu}\partial_{\mu}&f_0^q=p^{\mu}\partial_{\mu}(\exp(u_{\nu} p^{\nu}/T)+1)^{-1} \nonumber \\
		&=p^{\mu}f_0^q \tilde{f}_0^q \partial_{\mu}\left(\frac{u_{\nu}p^{\nu}}{T}\right) \nonumber \\
		&=\frac{p^{\mu}f_0^q \tilde{f}_0^q }{T}\left(u_{\nu}\partial_{\mu}p^{\nu}+p^{\nu}\partial_{\mu}u_{\nu}-u_{\nu}p^{\nu}\frac{\partial_{\mu} T}{T}\right).
		\label{qq}
\end{align}
Note that in the above equation, we have defined $\tilde{f}^0_q=1-f^0_q$. Also, making use of energy-momentum conservation (see Ref.~\cite{Chakraborty:2010fr}), Eq.~\eqref{qq} can further be simplified as	
\begin{align}
p^{\mu}\partial_{\mu}f_0^q=\frac{p^{\mu}f_0^q \tilde{f}_0^q}{T}&\bigg(u_{\nu}\partial_{\mu}p^{\nu}+p^{\nu}\partial_{\mu}u_{\nu}  \nonumber \\
& - u_{\nu}p^{\nu}(u^{\alpha}\partial_{\alpha}u_{\mu}-c_s^2 u_{\mu}\partial_{\alpha}u^{\alpha})\bigg).
		\label{qq1}
	\end{align}
For convenience, we solve Eq.~\eqref{qq1} by omitting the first term inside the bracket. It is important to note that, in the fluid rest frame, the medium effect is entirely captured by this first term, as it includes $p^0 \sim \epsilon_q$, which contains the temperature-dependent quasi-mass. By initially omitting the first term and addressing it separately, we simplify our calculations and clearly isolate the contribution arising purely from the medium effect. Therefore, dropping the first term,  Eq.~\eqref{qq1} reads 
	\begin{align}
		p^{\mu}&\partial_{\mu}f_0^q=\frac{p^{\mu}p^{\nu}f_0^q \tilde{f}_0^q }{T}\left(\partial_{\mu}u_{\nu}-u_{\nu}u^{\alpha}\partial_{\alpha}u_{\mu}+c_s^2 u_{\nu} u_{\mu}\partial_{\alpha}u^{\alpha}\right)\nonumber \\
		&=\frac{p^{\mu}p^{\nu}f_0^q \tilde{f}_0^q }{2T}\bigg(\nabla_{\mu}u_{\nu}+\nabla_{\nu}u_{\mu}-\frac{2}{3}\Delta_{\mu\nu}\partial_{\alpha}u^{\alpha}\nonumber \\
		 &\hspace{2.5cm}+\frac{2}{3}\Delta_{\mu\nu}\partial_{\alpha}u^{\alpha}+2c_s^2 u_{\nu} u_{\mu}\partial_{\alpha}u^{\alpha}\bigg).
		\label{qq2}
	\end{align}
With the further simplification of Eq.~\eqref{qq2}, we get the coefficients $\mathcal{A}$ and $\mathcal{C}$ (using Eqs.~\eqref{rtaBE} and~\eqref{phi}) as of the form 
	\begin{align}
		\mathcal{A}&=\frac{\tau_R\tilde{f}_0^q}{3T\epsilon_q}[(1-3c_s^2)\epsilon_q^2-m_q^2]~,
		\label{Aq}\\
		\mathcal{C}&=\frac{\tau_R\tilde{f}_0^q}{2T\epsilon_q}~.
	\end{align}
Now, we solve for the first term of Eq.~\eqref{qq1} which accounts for the medium effect, where we have the medium dependent mass $m_q \sim m_q(T)$. In the local rest frame of the fluid where the fluid velocity is defined as $u^{\mu}=(1,\vec{0})$, we get
	\begin{align}
		\frac{p^{\mu}f_0^q \tilde{f}_0^q }{T}u_{0}\partial_{\mu}p^{0}&=\frac{p^{\mu}f_0^q \tilde{f}_0^q }{T}\partial_{\mu}\epsilon_q \nonumber \\ 
		&= - c_s^2f_0^q \tilde{f}_0^q m_qm_q^{\prime} \partial_{\alpha}u^{\alpha}~.
		\label{qq3}
	\end{align} 
It is evident from the Eq.~\eqref{qq3} that incorporating the mean field effect alters the coefficient $\mathcal{A}$ and hence only the bulk viscosity. Therefore, the final form of the coefficient $\mathcal{A}$ is given as as the sum of the Eq.~\eqref{Aq} and Eq.~\eqref{qq3}. Hence,
	\begin{align}
		\mathcal{A}_q^{MF}=\frac{\tau_R\tilde{f}_0^q}{3T\epsilon_q}\left[(1-3c_s^2)\epsilon_q^2-m_q^2(T)+3c_s^2T^2\frac{dm_q^2(T)}{dT^2}\right]~,
		\label{AFq}
	\end{align}
and
	\begin{align}
		\mathcal{C}_{q}^{MF}=\mathcal{C}=\frac{\tau_R\tilde{f}_0^q}{2T\epsilon_q}~,
		\label{CFq}
	\end{align}
where the subscript($MF$) stands for mean-field. Hence, the shear and the bulk viscosity for the quasiparticle quarks, using Eqs.~\eqref{zeta},~\eqref{eta},~\eqref{AFq} and~\eqref{CFq}, and following the Landau frame condition (discussed in details in next subsection)
	\begin{align}
		\eta_{q/\bar{q}}&=\frac{\tau_R}{15T}\int \frac{d^3\textbf{\textit{p}}}{(2\pi)^3}\frac{\textbf{\textit{p}}^4}{\epsilon_q^2}f^{0}_q(\epsilon_q) \tilde{f}^0_q(\epsilon_q)~.
		\label{etaq}\\
		\zeta_{q/\bar{q}}&=\frac{\tau_R}{9T}\int  \frac{d^3 \textbf{\textit{p}}}{(2\pi)^3}\frac{f^{0}_q(\epsilon_q)\tilde{f}^0_q(\epsilon_q)}{\epsilon_q^2} \nonumber \\ 
		&\times\bigg[\left(1-3c_s^2\right)\epsilon_q^2-m_q^2(T)+3c_s^2T^2\frac{dm_q^2(T)}{dT^2}\bigg]^2~.
		\label{zetaq}
	\end{align}
Here, the Fermi blocking factor $\tilde{f}_q^0$ is a direct consequence of the Eq.~\eqref{qq}. Likewise, for the gluonic case, we will see the Bose enhancement factor $\tilde{f}_{gl}^0$ as discussed in the next subsection. In the limit of zero chemical potential, the shear and bulk viscosities from quarks and antiquarks are identical. Accordingly, the subscript $q/\bar{q}$ in Eqs.~\eqref{etaq} and~\eqref{zetaq} denotes quarks and antiquarks, respectively.
	\subsection{Shear and Bulk viscosity of Gribov-modified gluons}
In this section, the transport coefficients of the Gribov-modified gluonic plasma have been formulated. Note that there are two quasi-poles, given by Eq.~\eqref{epsilon1} and Eq.~\eqref{epsilon2}. For this set of energies, we will have two different forms of $\mathcal{A}$ and $\mathcal{C}$ corresponding to each of them, determining the form of $\zeta$ and $\eta$ for the gluonic plasma.
	
While evaluating the coefficients, the method adopted in the previous section will again be applied. It is important to realise that difference that occurs primarily due to the first term in the Eq.~\eqref{qq} (as a result of different forms of the energy density between the quasiparticle quarks and gluons), and also here, the presence of two different poles, gluonic plasma consists of two different form of $\mathcal{A}$ (corresponding to the Eq.~\eqref{epsilon1} and Eq.~\eqref{epsilon2}). The same holds for the coefficient $\mathcal{C}$. Without making much of the efforts, it is clear that 
	\begin{align}
		\mathcal{C}_{gl}^{MF}(\epsilon_1)=\frac{\tau_R\tilde{f}_{gl}^0}{2T\epsilon_1}~,
\\
		\mathcal{C}_{gl}^{MF}(\epsilon_2)=\frac{\tau_R\tilde{f}_{gl}^0}{2T\epsilon_2}~,
	\end{align}
where $\epsilon_1$ and $\epsilon_2$ are the energies defined in Eqs.~\eqref{epsilon1} and~\eqref{epsilon2} and $\tilde{f}_0^{gl}=1+f_0^{gl}$ with $f_0^{gl}$ being the local equilibrium distribution function for gluons, defined in Eq.~\eqref{distributionfngluon}. Therefore, using Eq.~\eqref{eta}, the coefficient of shear viscosity (corresponding to the two different coefficients $\mathcal{C}(\epsilon_1)$ and $\mathcal{C}(\epsilon_2)$ is given by
	\begin{align}\label{etagl1}
		\eta_{gl}^1=\frac{\tau_R}{15T}\displaystyle\int \frac{d^3\textbf{\textit{p}}}{(2\pi)^3}\frac{\textbf{\textit{p}}^4}{\epsilon_1^2}f_{gl}^{0}(\epsilon_1)\tilde{f}_{0gl}^{0}(\epsilon_1)~.
	\end{align}
	In the similar manner, we get
	\begin{align}\label{etagl2}
		\eta_{gl}^2=\frac{\tau_R}{15T}\displaystyle\int \frac{d^3\textbf{\textit{p}}}{(2\pi)^3}\frac{\textbf{\textit{p}}^4}{\epsilon_2^2}f_{gl}^{0}(\epsilon_2)\tilde{f}_{gl}^{0}(\epsilon_2)~.
	\end{align} 
The final expression of the shear viscosity of the Gribov gluon is
	\begin{align}\label{etagluon}
		\eta_{gl}=\eta_{gl}^1+\eta_{gl}^2~.
	\end{align}
	\begin{figure*}
	\centering
	\includegraphics[scale=0.6]{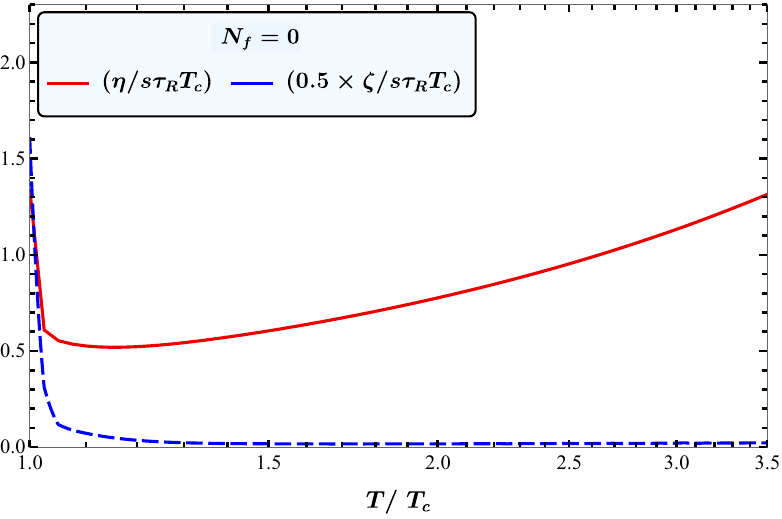}
	\hspace{1cm}
	\includegraphics[width=8.25cm]{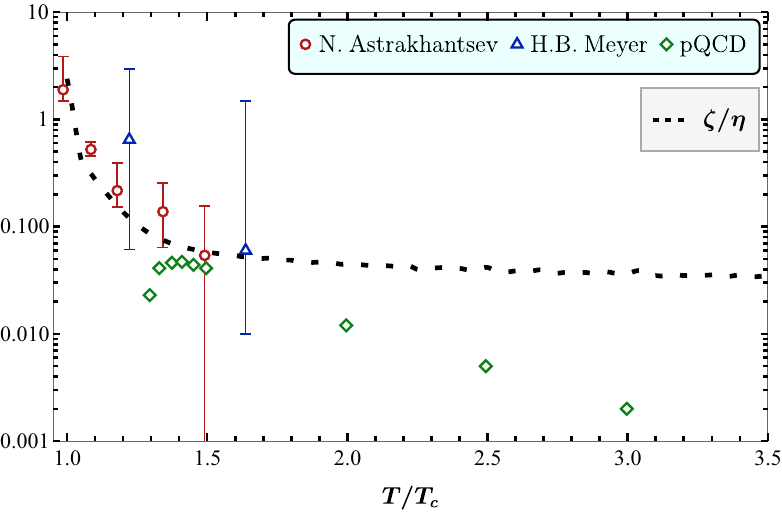}
	\caption[eta gluon]{On left, temperature dependence of the scaled shear viscosity from Eq.~\eqref{etagluon} and the bulk viscosity from Eq.~\eqref{zetagluon}) are plotted. The scaling is done with $s\tau_R T_c$, where $s$ is the entropy density of Gribov-modified gluonic plasma. The scaled temperarure runs from $1 \le T/T_c \le 3.5$. The plot on the right shows the temperature dependence of $\zeta/\eta$ of gluonic plasma as a function of $T/T_c$. Lattice data has been taken from Refs.~\cite{Meyer:2007dy, Meyer:2007ic} (blue triangle), Refs.~\cite{Astrakhantsev:2018oue, Astrakhantsev:2017nrs} (red circle) and Refs.~\cite{Arnold:2006fz, Arnold:2003zc} (green diamond, pQCD).}
	\label{fig:transportnf0}
	\end{figure*}
In Ref.~\cite{Peshier:2005pp}, the shear viscosity of the gluonic plasma has been explored in quasiparticle approximation with Lorentz spectral function. The quasiparticle gluonic mass $M$ and the spectral width $\Gamma$ has been parameterized with the running coupling  which in turn is fitted using the dynamical quasiparticle entropy density with the lattice QCD. In this approach, 1-loop Dyson-Schwinger equation is used to obtained the form of the resummed propagator.
The study reveal that the gluon plasma show liquid-like behavior, supported by a very low and nearly constant shear viscosity-to-entropy ratio ($\eta/s\sim 0.2$) in the temperatures range $1.05 \le T/T_c \le 4$. However, in the present approach, we observe an increasing trend in the scaled shear viscosity with the additional scaled parameter $\tau_R T_c$.

Now, while formulating the bulk viscosity for gluons, the coefficients $\mathcal{A}_1$(corresponding to $\epsilon_1$) and $\mathcal{A}_2$ (corresponding to $\epsilon_2$) are needed to be determined. The uniqueness of these coefficients is given by $\mathcal{A}_1^{par}$ and $\mathcal{A}_2^{par}$, where $``par"$ signifies the particular solution of the coefficient, which are realized for as long as the Landau frame condition is satisfied (see Ref.~\cite{Chakraborty:2010fr}), the detail of which could also be found in the Appendix[\ref{appen:gribovgluon}]. Therefore, solving the Boltzmann equation (Eq.~\eqref{qq}) with the only change in the form of the distribution function, we get 	
\begin{equation}
		\begin{split}
			\mathcal{A}^{par}_1&=\frac{\tau_R\tilde{f}_0^{gl}(\epsilon_1)}{3T\epsilon_1}
			\bigg[\textbf{\textit{p}}^2-3c_s^2\epsilon_1^2  \\ 
			&+\frac{3c_s^2T}{2}\bigg(m_gm_g^{\prime}+\frac{m_g^3m_g^{\prime}-4\gamma_G^3\gamma_G^{\prime}}{\sqrt{m_g^4-4\gamma_G^4}}\bigg)\bigg]~,
			\label{A1}
		\end{split}
\end{equation}
In the same way, doing the calculation corresponding to $\epsilon_2$ (Eq.~\eqref{epsilon2}), we get
	\begin{equation}
		\begin{split}
			\mathcal{A}^{par}_2&=\frac{\tau_R\tilde{f}_0^{gl}(\epsilon_2)}{3T\epsilon_2}\bigg[\textbf{\textit{p}}^2-3c_s^2\epsilon_2^2\\
			&+\frac{3c_s^2T}{2}\bigg(m_gm_g^{\prime}+\frac{4\gamma_G^3\gamma_G^{\prime}-m_g^3m_g^{\prime}}{\sqrt{m_g^4-4\gamma_G^4}}\bigg)\bigg]~.
			\label{A2}
		\end{split}
	\end{equation}
In the above equations, $c_s^2$ is the square of speed of sound (for gluonic medium), which is discussed in section~\ref{sec:vssq} in detail. Now, considering the mean field effect along with the Landau frame condition, the final expression for Bulk viscosity $\zeta$ for Gribov-modified gluons takes the form (corresponding to $\mathcal{A}^{par}_1$), using Eqs.~\eqref{zetagluon1LL} and~\eqref{A1}	
	\begin{align}\label{zetagl1}
		&\zeta _{gl}^1=\frac{\tau_R}{9T}\int \frac{d^3\textbf{\textit{p}}}{(2\pi)^3}\frac{f_0^{gl}(\epsilon_1)\tilde{f}_0^{gl}(\epsilon_1)}{\epsilon_1^2} \nonumber \\
		&\ \ \ \times \bigg[\textbf{\textit{p}}^2-3c_s^2\epsilon_1^2+\frac{3c_s^2T}{2}\bigg(m_gm_g^\prime+\frac{m_g^3m_g^\prime-4\gamma_G^
			3\gamma_G^\prime}{\sqrt{m_g^4-4\gamma_G^4}}\bigg)\bigg]^2 \!.\hspace{-.1cm}
	\end{align}
Similarly, the bulk viscosity expression, (corresponding to $\mathcal{A}^{par}_2$), using Eq.~\eqref{zetagluon2LL} and Eq.~\eqref{A2} becomes
	\begin{align}\label{zetagl2}
		&\zeta _{gl}^2=\frac{\tau_R}{9T}\int \frac{d^3\textbf{\textit{p}}}{(2\pi)^3}\frac{f_0^{gl}(\epsilon_2)\tilde{f}_0^{gl}(\epsilon_2) }{\epsilon_2^2}\nn \\ 
		&\ \ \ \times \bigg[\textbf{\textit{p}}^2-3c_s^2\epsilon_2^2+\frac{3c_s^2T}{2}\bigg(m_gm_g^\prime+\frac{4\gamma_G^3\gamma_G^\prime-m_g^3m_g^\prime}{\sqrt{m_g^4-4\gamma_G^4}}\bigg)\bigg]^2\!.\hspace{-.1cm}
	\end{align}
The final expression of the bulk viscosity of the Gribov-modified gluon is
\begin{align}\label{zetagluon}
		\zeta_{gl}=\zeta_{gl}^1+\zeta_{gl}^2.
	\end{align}
A similar analysis of the shear and bulk viscosities for Gribov gluons has been conducted previously in both the covariant gauge (Ref.~\cite{Jaiswal:2020qmj}) and the Coulomb gauge (Ref.~\cite{Begun:2016lgx}), utilizing the simplest form of the Gribov-modified gluon propagator, which includes only the $\gamma_G$ term. In the current study, the transport coefficients exhibit an enhancement across the entire range of scaled temperatures, likely due to the inclusion of the dynamical gluon mass $m_g$.

The perturbative QCD (pQCD) analysis on the other hand, carried out by P.B. Arnold \textit{et al.} of the shear and the bulk viscosity in the Ref.~\cite{Arnold:2003zc} and Ref.~\cite{Arnold:2006fz}, respectively. Their study employed the next-to-leading-log (NLL) expansion, yielding the following expressions for the transport coefficients:
\begin{align}
	\eta_{\rm NLL}= \frac{T^3}{g^4}\left(\frac{\eta_1}{\ln(\rho_1/M_d)}\right); {\hspace{0.2cm}} \zeta_{\rm NLL}=\left(\frac{\mathcal{A}\alpha_s^2T^3}{ ln(\rho_2/M_d)}\right)~,
\end{align}
where the Debye mass $M_d/T=g$ and $\alpha_s=g^2/4\pi$ is the coupling. For the purely gluonic case, these parameters takes the values $\mathcal{A}=0.443$, $\eta_1=27.126$, $\rho_1/T=2.765$ and $\rho_2/T=7.14$. The decrease of $\zeta/\eta$ value below $T\lesssim 1.4T_c$ observed  in the right panel of Fig. (\ref{fig:transportnf0}), marked by the pQCD result (green diamond) highlights the limitations of applying pQCD in the region near $T \rightarrow T_c$. Relying on a logarithmic expansion, pQCD calculation is relevant when the logarithmic term in the denominator is large (signifying lower value of the running coupling, $M_d \sim gT$). However, as the running coupling increases closer to the critical temperature, logarithmic expansion begins to break down. Interestingly, the decrease also indicate that the logarithmic expansion for $\eta$ breaks down at slightly smaller running coupling value compared to $\zeta$. Noting that the perturbative approach assumes a small running coupling, a condition that is no longer valid as $T \rightarrow T_c$ further reduces its applicability in this region.

It is interesting to see that the influence of medium effects has a more pronounced impact on the bulk viscosity than the shear viscosity.Thus, the impact of medium effects on transport properties is more evident in the bulk viscosity. The ratio ($\zeta/\eta$) is also parameterized in terms of the speed of sound square, allowing for a comprehensive investigation of the behaviour of the QGP medium. Hence, a physical significance is associated with this ratio $\zeta/\eta$. With shear and bulk viscosities in hand, we see the variation of the ratio $\zeta/\eta$ as a function of scaled temperaturein the Fig.~(\ref{fig:transportnf0}). It can be observed from the plot of $\zeta/\eta$ against the scaled temperature that around $T/T_c \sim 1.4$, our results are in close agreement with perturbative predictions. However, as $T/T_c$ increases, the expected restoration of conformal invariance is suppressed. This suppression is likely due to the reduced shear viscosity, which results from enhanced interactions (and thus increased cross-sections) in the Gribov-modified gluonic plasma.
\subsection{Shear and Bulk viscosity of QGP}
\begin{figure}
		\centering
		\includegraphics[scale=0.62]{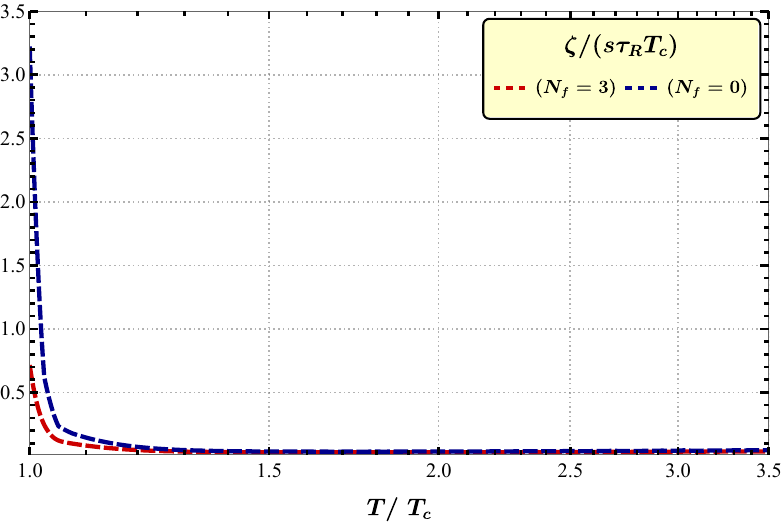}
		\includegraphics[scale=.63]{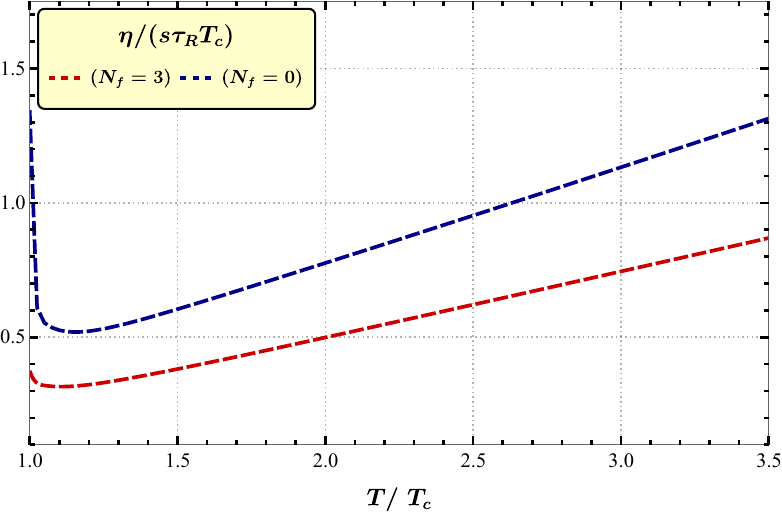}
		\caption[zeta qgp]{The bulk viscosity (plot n the top) of the medium QGP (red dotted) and Gribov-modified gluon (blue dotted), scaled with $s\tau_RT_c$ plotted against the scaled temperature in the range ($1 \le T/T_c \le 3.5$). Scaled shear viscsity is also shown in the bottom plot.}
		\label{fig:etazetafinal}
	\end{figure}

As stated earlier, we are considering the QGP medium consists of the quasiparticle quarks and gluons modified by the Gribov prescription. Moreover, the interaction between the quasiparticle quarks and gluons are incoded in the quasi mass of the quarks $m_q$. The temperature dependence of the transport coefficients (shear and bulk viscosities) of the medium QGP is given as a sum of the transport coefficients from the quasiparicle quarks and the Gribov modified gluons, i.e. from  Eqs.~\eqref{etagluon},~\eqref{zetagluon},~\eqref{etaq} and~\eqref{zetaq} as
\begin{align}
		\zeta_{QGP}&=\zeta_{gl}+\zeta_{q/\bar{q}}~, {\hspace{0.1cm}} {\rm and}
\label{zetaqgp}\\
		\eta_{QGP}&=\eta_{gl}+\eta_{q/\bar{q}}~.
		\label{etaqgp}
	\end{align}
The variation of Eqs.~\eqref{zetaqgp} and~\eqref{etaqgp} with the scaled temperature $T/T_c$ has been shown in Fig.~(\ref{fig:etazetafinal}). While dealing with the transport coefficients of the QGP medium, particularly bulk viscosity, we make use of the velocity of sound square of the QGP medium, as discussed in detail in section~\ref{sec:vssq}.
The transport coefficients, such as shear and bulk viscosities, have been extensively studied within the quasiparticle approximation. In Refs.~\cite{Mykhaylova:2020pfk, Mykhaylova:2019wci}, a detailed investigation was carried out on the impact of quasiparticle quarks on the shear viscosity of QGP, alongside the analysis of bulk viscosity based on the same formalism. The study explores the effects of different quark flavors on shear viscosity, with the quasi-gluon and quasi-quark masses parameterized in terms of the running coupling, which is fitted using lattice QCD equation of state.\newline
By this approach, the conformal limit is attained more quickly as compared to the present approach suggesting a greater interaction among the medium particles and hence more near-to ideal fluid behaviour.
\section{Bulk to Shear viscosity analysis}\label{bulktoshear}
In this section, we analyze the ratio of bulk to shear viscosity for the gluonic plasma, as well as for the medium QGP. A low value of $\zeta/\eta$ implies that shear viscosity dominates over bulk viscosity, reflecting a medium more responsive to shape deformations than volume changes. In contrast, a higher $\zeta/\eta$ ratio, particularly near the transition region, indicates that volume-changing processes are more significant.
	
At high temperatures, the perturbative calculation predicts $\zeta/\eta \propto (\Delta c_s^2)^2$~\cite{Jeon:1995zm, Florkowski:2015lra} where we have used the notation $\Delta c_s^2= (1/3-c_s^2)$. The same trend is also observed for an interacting photon gas in Ref.~\cite{Weinberg:1971mx} and in scalar field theory in Ref.~\cite{Jeon:1995zm}. Interestingly, near the transition region from hadronic to QGP phase, calculation from gauge-gravity duality predicts the linear dependence on $\Delta c_s^2$~\cite{Buchel:2005cv, Benincasa:2005iv}. 
	
\begin{figure}
		\centering
		\includegraphics[scale=0.6]{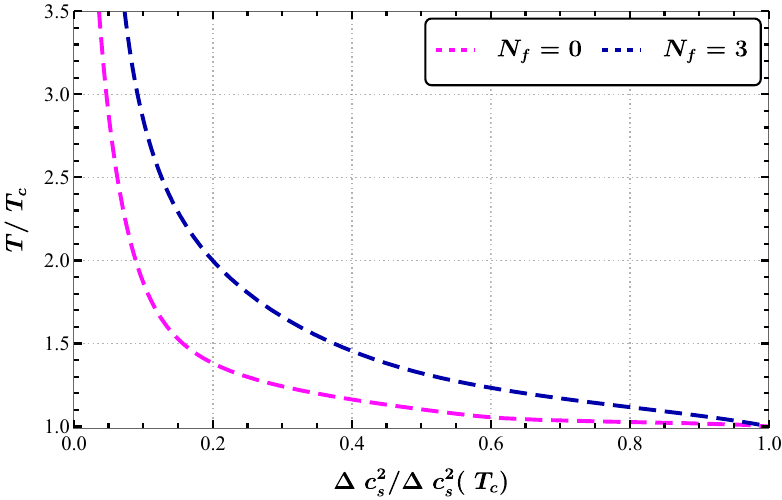}
		\caption{The scaled temperature $T/T_c$ against the conformality measure, normalized by its value at $T_c$.}
		\label{fig:delcssqvst}
	\end{figure}

We have compared our finding of ratio $\zeta/\eta$ to that of linear and quadratic dependence on $\Delta c_s^2$, whose forms are given as	
\begin{align}
		{\rm Linear}: \frac{\zeta}{\eta}=\alpha (\Delta c_s^2)+\beta, {\rm and}  
		\label{linear}\\
		{\rm Quadratic}: \frac{\zeta}{\eta}=\gamma(\Delta c_s^2)^2+\delta~,
		\label{quad}
\end{align}
where $\alpha$, $\beta$, $\gamma$ and $\delta$ are the fit parameters. Similar analysis has also been explored in Ref.~\cite{Mykhaylova:2020pfk} in pure Yang-Mills as well as medium QGP in the flavor dependence of transport coefficients in a quasi-particle approach.
	\begin{figure}
		\centering
		\includegraphics[width=8.25cm]{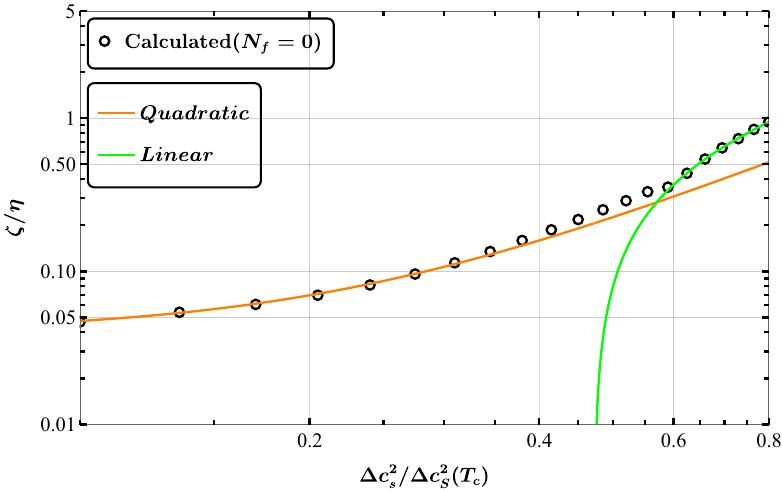}
		\includegraphics[width=8.25cm]{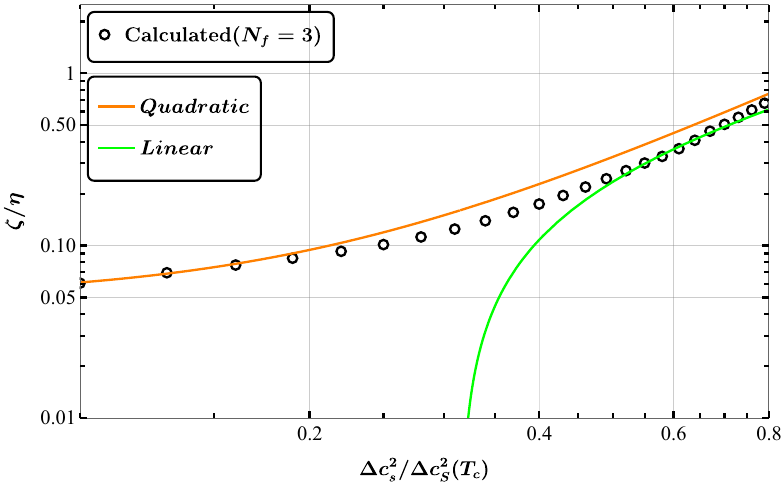}
		\caption{Bulk to shear viscosity ratio of gluonic plasma [using Eq.~\eqref{zetagluon} and~\eqref{etagluon}]  against conformality measure $\Delta c_s^2$. The fit parameters for $N_f=0$, corresponding to linear dependence are $\alpha=2.88$ and $\beta=-1.36$ while for quadratic dependence, $\gamma=0.745$ and $\delta=0.04$. For $N_f=3$ (plot at the bottom) [using Eqs.~\eqref{zetaqgp} and~\eqref{etaqgp}], the fit parameters are $\alpha=1.27,\beta=-0.4, \gamma=1.11, \delta=0.05$.}
		\label{fig:zetabyetanf0/3sp}
	\end{figure}
In Fig.~(\ref{fig:zetabyetanf0/3sp}), the variation of ratio($\zeta/\eta$) against $\Delta c_s^2$ normalized by its value at a transition temperature $T_c$ of Gribov-modified gluonic plasma has been depicted. We notice a good agreement with the quadratic scaling in the high-temperature limit, while for the intermediate temperature range, linear scaling is a good match, which is in line with the previous findings based on the behaviour of strongly interacting mediums. From Fig.~(\ref{fig:delcssqvst}),  it can be seen that the shift from the linear scaling to quadratic scaling (${\rm for} \hspace{0.1cm} N_f=0$) occurs around $T/T_c \sim 1.2$.
		
Another approach which describes the pure Yang-Mills plasma based on the Gribov–Zwanziger quantization (based on a constant Gribov parameter, $\gamma_G$), leads to the ratio $\zeta/\eta$ linearly proportional to the quantity $\Delta c_s^2$~\cite{Florkowski:2015dmm}. In the high-temperature limit
	\begin{align}
	\zeta/\eta=\mathcal{X}_{GZ}\left(\frac{1}{3}-c_s^2\right)+\cdots(T \gg \gamma_G)~,    
	\end{align}
where $\mathcal{X}_{GZ}=5/2$. However, in the present analysis, we have considered the temperature dependence Gribov parameter $\gamma_G(T)$, whose form is given by Eq.~\eqref{gribovparameter}. For BE plasma (with effective mass, $m_\text{eff}$) at high temperatures, again, from Ref.~\cite{Florkowski:2015dmm}, it was found that, at high temperatures
	\begin{align}
		\frac{\zeta}{\eta}=\mathcal{X}_{BE}\left(\frac{1}{3}-c_s^2\right)^{3/2}+\cdots(T \gg m_\text{eff})~,   
	\end{align}
	with $\mathcal{X}_{BE}\sim 5.81$. Also, at the low temperature expansion, where $c_s^2\rightarrow 0$, corresponds to $\mathcal{X}_{GZ}=5$ while $\mathcal{X}_{BE}=2$.
	
A similar analysis is done also for $N_f=3$ as shown in Fig.~(\ref{fig:zetabyetanf0/3sp}). The transition from linear dependence to quadratic the dependence is quite evident and occurs near $T/T_c \sim 1.6$.			
\section{Summary and Outlook}\label{sec:summary}
In this study, we have explored the transport coefficients namely, shear and bulk viscosity of the medium QGP, consisting of the Gribov-modified gluons and quasiparticle quarks. The Gribov idea, known for its efficacy in capturing non-perturbative effects, has been employed to analyze the thermodynamics along with the transport coefficients of gluonic plasma.
	
For the gluonic plasma, we identify two key variables: the Gribov parameter $\gamma_G$ and the dynamical mass $m_g$. We first apply the gap equation, allowing us to express the dynamical mass in terms of the Gribov parameter ($m_g\sim m_g(\gamma_G)$), numerically. Subsequently, we used the lattice equation of state for quenched QCD, taken from Ref.~\cite{Borsanyi:2012ve}, to fit the temperature dependence of the Gribov parameter, which in turn determines the temperature dependence of the dynamical mass.
 The resulting variation with the temperature of these variables are illustrated in Fig.~(\ref{fig:mgandgammag}).
 For the quark sector, we adopt the quasiparticle approach, where the energy is characterized by $\epsilon_q = (\textbf{\textit{p}}^2 + m_q^2)^{1/2}$. Here, the quasiparticle mass of the quark, denoted as $m_q$, is determined using lattice data for the entropy density from Ref.~\cite{Borsanyi:2010cj}. Notably, the quark-gluon interactions are encapsulated within the quasiparticle mass, $m_q$. The temperature dependence of scaled (by temperature $T$) $m_q$ is shown in Fig.~(\ref{fig:mq}).

The plot on the left in Fig.~(\ref{fig:transportnf0}) illustrates the variation of shear and bulk viscosities, scaled by $s \tau_R T_c$
, for the gluonic plasma modified by the Gribov prescription, as a function of the scaled temperature with $T_c$ set at 260 MeV. 
 The observed behavior of both transport coefficients matches well with theoretical expectations and previous studies. A sharp increase in the bulk viscosity is evident near the critical temperature, followed by a rapid decrease as the temperature rises, signifying that the system approaches the conformal limit at higher temperatures. This trend is consistent with the established understanding. Similarly, the scaled shear viscosity exhibits a minimum near $T_c$, a feature commonly anticipated in such systems, after which it increases with  temperature. The plot on the right shows the ratio $\zeta/\eta$ obtained which shows a good agreement with the corresponding quantity derived from existing lattice data from Refs.~\cite{Meyer:2007dy, Meyer:2007ic} and also from Refs.~\cite{Astrakhantsev:2018oue, Astrakhantsev:2017nrs} and from the Refs.~\cite{Arnold:2006fz, Arnold:2003zc}. It is also worth noting that the lattice findings are stabbed with huge uncertainties near the transition region.  Above $T/T_c=1.2$, the ratio are in good agreement with the lattice finding. We also notice a bit of gradual restoration of the conformal limit as the temperature goes up, in the Gribov-modified approach.
 	
The Fig.~(\ref{fig:etazetafinal}) presents the comparison of scaled bulk viscosity (on top) for both the gluonic plasma and the medium QGP. The interaction between quarks and gluons is encoded in the quasiparticle mass of the quarks, where three degenerate quark flavors (u, d, and s) with equal quasi-mass are considered. The plot reveals a significant suppression in the scaled bulk viscosity after incorporating the effect of quasiparticle quarks, observed consistently across the entire temperature range.
While the plot at the bottom, we presents the variation of the scaled shear viscosity $\eta/s\tau_RT_c$ for the same. A minimum in the scaled shear viscosity is observed near the critical temperature ($T_c$), a trend that holds for both $N_f=0$ and $N_f=3$ across the relevant temperature range ($1 \le T/T_c \le 3.5$). The blue dashed curve represents the variation of shear viscosity for the gluonic plasma, while the red dashed curve corresponds to the medium QGP. As temperature increases, the shear viscosity rises, a behavior that can be attributed to the growing influence of particle collisions and scattering processes. These interactions enhance the transfer of momentum between different regions of the fluid, leading to a greater resistance to shear deformation, thus elevating the shear viscosity. This trend aligns with the expected results from both kinetic theory and lattice-based studies, further validating the reliability of the theoretical framework employed in this analysis.
	
In Fig.~(\ref{fig:zetabyetanf0/3sp}), we have examined the ratio of bulk to shear viscosity for both, gluonic plasma and the medium QGP. The results indicate a transition from linear to quadratic scaling as temperature increases. For the gluonic plasma(top plot), this transition occurs at approximately at $T/T_c\sim 1.2$, while for the medium with quark contributions (at the bottom), the shift is observed around $T/T_c\sim 1.6$. The transition temperature can be inferred from Fig.~(\ref{fig:delcssqvst}), which provides insight into the point where the scaling behavior changes.
	
The present approach to exploring the transport coefficients of QGP is not intended as a precise phenomenological model for investigating these coefficients. Rather, it serves as a preliminary framework for incorporating non-perturbative effects via the modified Gribov prescription by incorporating the condensate related parameter $m_g$, providing valuable insights into the transport properties of the hot QGP medium. This is particularly important near the transition region, where significant changes in the system’s behavior occur. Future research could focus on an in-depth study of the relaxation time ($\tau_R$) and its implications. Additionally, the flavor dependence of the system, especially in the context of recovering the conformal limit, could be explored in detail. By incorporating these refinements, we can establish a more comprehensive foundation for validating our results, particularly through comparison with lattice QCD calculations of scaled transport coefficients available in Refs.~\cite{Nakamura:2004sy, Sakai:2007cm, Astrakhantsev:2018oue, Meyer:2007dy}. In addition,  we adopt the refined Gribov prescription, where two distinct mass terms, $M_g$ and $m_g$, arise in the gluon propagator which are determined individually solving the gap equations obtained by minimizing the Gribov action with respect to these variables. However, in the present analysis, we have chosen to include only the dynamical mass term $m_g$ and have neglected any contributions from the $M_g$ term. This simplification represents a limitation of the current study, as the inclusion of $M_g$ would significantly complicate the analysis. The $M_g$ term, which emerges from the contributions of auxiliary fields introduced in the refined Gribov-Zwanziger framework, plays a key role in capturing additional non-perturbative effects in the gluon sector. Incorporating the $M_g$ term in future studies would provide a more comprehensive description of the system, at the cost of increased complexity. This step would require a more thorough treatment of the gap equations and a careful consideration of the interplay between $M_g$ and $m_g$. Given these challenges, this extension is deferred for future work, where a more detailed investigation into the combined effects of both mass terms could yield further insights into the non-perturbative dynamics of the QGP.	

	\section{Acknowledgements}
	The author acknowledges the fruitful discussion with Ritesh Ghosh and Sumit. N.H. is supported in part by the SERB-MATRICS under Grant No. MTR/2021/000939.
	\appendix
	\section{Landau frame condition: Gribov gluons}\label{appen:gribovgluon}
	Here, we see the effect of the Landau frame condition along with the mean-field effect on the transport coefficients of Gribov-modified gluonic plasma. This is achieved by following method outlined in Ref.~\cite{Chakraborty:2010fr}.
	
	For a small deviation away from the equilibrium, the change in distribution function is
	\begin{align}
		f(x,p)=f^{eq}(\epsilon_{0})+\delta f(x,p)~,
	\end{align}
	where $\epsilon_{0}$ correspond to the equilibrium energy. In order to capture the non-equilibrium deviation in the energy, we define $\epsilon$, such that
	\begin{align}
		\epsilon=\epsilon_0+\delta \epsilon.
	\end{align}
	The distribution function considering the off-equilibrium shift is
	\begin{align}
		f(x,p)=f^{eq}(\epsilon)+\delta \tilde{f}(x,p)~,
	\end{align}
	where
	\begin{align}
		\delta \tilde{f}(x,p)=\delta f(x,p)-\frac{\partial f^{eq}(\epsilon)}{\partial \epsilon}\delta \epsilon~.
		\label{deltaftilde}
	\end{align}
	Now, for the out of equilibrium case, we defined
	\begin{align}
		\Delta T^{ij}=\int \frac{d^3\textbf{\textit{p}}}{(2\pi)^3}\frac{\textbf{\textit{p}}^i\textbf{\textit{p}}^j}{\epsilon}\delta \tilde{f}~.
		\label{deltatij}
	\end{align}
	Using Eq.~\eqref{deltaftilde} in Eq.~\eqref{deltatij} and a bit of simplification, the shift in energy density is
	\begin{align}
		\Delta T^{00} &=\int \frac{d^3\textbf{\textit{p}}}{(2\pi)^3} \left(\epsilon \delta \tilde{f}+\frac{\partial f^{eq}(\epsilon)}{\partial \epsilon}\epsilon \delta \epsilon\right) \nn \\
		&= \int \frac{d^3\textbf{\textit{p}}}{(2\pi)^3} \left(\epsilon \delta \tilde{f}+\frac{f^{eq}(\epsilon)\tilde{f}^{eq}(\epsilon)}{T}\epsilon \delta \epsilon\right) .
		\label{deltaT00}
	\end{align}
	At $\epsilon=\epsilon_1$ [as given in Eq.~\eqref{epsilon1}], Eq.~\eqref{deltaT00} becomes
	\begin{align}
		\Delta T^{00}&=\int \frac{d^3\textbf{\textit{p}}}{(2\pi)^3}\Bigg[\epsilon_1 \delta \tilde{f} \nn \\
		& +\frac{f^{eq}\tilde{f}^{eq}}{2T}\left(m_gm_g^{\prime}-\frac{m_g^3m_g^{\prime}-4\gamma_G^3\gamma_G^{\prime}}{\sqrt{m_g^4-4\gamma_G^4}}\right)\delta T \Bigg]~,
		\label{deltaT00ep1}
	\end{align}
	where, for obvious reason, $f^{eq}\tilde{f}^{eq}=f^{eq}(\epsilon_1)\tilde{f}^{eq}(\epsilon_1)$. Now, for the deviation away from equilibrium, the distribution function takes the form
	\begin{align}
		\delta f=-f^{eq}(\epsilon_1) \tilde{f}^{eq}(\epsilon_1)\left(\frac{\delta \epsilon_1}{T}-\frac{\epsilon_1}{T^2}\delta T\right)~.
		\label{deltaf}
	\end{align}
	Using Eq.~\eqref{deltaf} in Eq.~\eqref{deltaftilde}, $\delta T$ for $\epsilon=\epsilon_1$ simplifies as
	\begin{align}
		\delta T=\frac{T^2}{f^{eq}(\epsilon_1)\tilde{f}^{eq}(\epsilon_1)\epsilon_1}\delta \tilde{f}
		\label{deltat}.
	\end{align}
	Now, we substitute Eq.~\eqref{deltat} back in Eq.~\eqref{deltaT00ep1}
	\begin{align}
		\Delta T^{00}_{\epsilon=\epsilon_1}&=\Delta \epsilon_1=\int \frac{d^3 \textbf{\textit{p}}}{(2\pi)^3} \frac{1}{\epsilon_1} \nn \\
		&\times\left[\epsilon_1^2-\frac{T}{2}\left(m_gm_g^\prime+\frac{m_g^3m_g^\prime-4\gamma_G^3\gamma_G^\prime}{\sqrt{m_g^4-4\gamma_G^4}}\right)\right]\delta \tilde{f}.
		\label{a}
	\end{align}
	Following the argument of~\cite{Chakraborty:2010fr}, the more generalized form of out-of-equilibrium energy-momentum tensor, for $\epsilon=\epsilon_1$, is
	\begin{align}
		\Delta T^{\mu\nu}&=	\int\frac{d^3\textbf{\textit{p}}}{(2\pi)^3}\frac{1}{\epsilon_1}  \Bigg[p^{\mu}p^{\nu}\nonumber \\
		& -u^{\mu}u^{\nu}\frac{T}{2}\left(m_gm_g^\prime+\frac{m_g^3m_g^\prime-4\gamma_G^3\gamma_G^\prime}{\sqrt{m_g^4-4\gamma_G^4}} \right)\Bigg] \delta \tilde{f}.
	\end{align}
	In the context of hydrodynamics, the Landau frame condition is often used to analyze the behaviour of fluctuations in a fluid. When applied to the study of viscosity, it specifically affects the behaviour of bulk viscosity $\zeta$ and not shear viscosity $\eta$. As it is a well-established fact that the shear viscosity is related to the dissipation of momentum due to the relative motion of adjacent fluid layers, it does not directly involve the dispersion relation or the speed of sound, $c_s^2$. Therefore, the Landau frame condition primarily influences the behavior of bulk viscosity, which is associated with energy dissipation due to fluid volume changes.
	
	One of the subtleties in $\mathcal{A}$ is an infinite number of solutions could be generated by shifting
	\begin{eqnarray}
	\mathcal{A}(\epsilon_1)\rightarrow \mathcal{A}'(\epsilon_1)=\mathcal{A}(\epsilon_1)-a-b\epsilon_1~,
	\end{eqnarray}
	where for zero chemical potential, $a=0$ since it is connected with the particle number conservation.
	
	For a particular solution, say $\mathcal{A}^{par}_1$, we take
	\begin{align}
		\mathcal{A}=\mathcal{A}^{par}_1-b\epsilon_1~.
		\label{apar}
	\end{align}
	Using the Landau frame condition, $u_{\mu}\Delta T^{\mn}=0$, we get
	\begin{align}
		\int &\frac{d^3\textbf{\textit{p}}}{(2\pi)^3}\frac{f_g^{eq}(\epsilon_1)\tilde{f}_g^{eq}(\epsilon_1)}{\epsilon_1} (\mathcal{A}^{par}_1-b\epsilon_1) \nonumber \\
		&\times \left[\epsilon_1^2-\frac{T}{2}\left( m_gm_g^\prime+\frac{m_g^3m_g^\prime-4\gamma_G^3\gamma_G^\prime}{\sqrt{m_g^4-4\gamma_G^4}}\right)\right] =0 .
	\end{align}
	Making use of the thermodynamic relation
	$(d\mathcal{P}/dT)=(d\mathcal{P}/d\epsilon)(d\epsilon/dT)=c_s^2(d\epsilon/dT)$, we get 
	\begin{align}
		b&=\frac{c_s^2}{T^2s}\int \frac{d^3\textbf{\textit{p}}}{(2\pi)^3}\frac{f_g^{eq}(\epsilon_1)\tilde{f}_g^{eq}(\epsilon_1)}{\epsilon_1} \nn \\
		&\times\left[\epsilon_1^2-\frac{T}{2}\left( m_gm_g^\prime+\frac{m_g^3m_g^\prime-4\gamma_G^3\gamma_G^\prime}{\sqrt{m_g^4-4\gamma_G^4}}\right)\right]\mathcal{A}^{par}_1,
		\label{b}
	\end{align}
	where, 
	\begin{align}
		s=\frac{1}{3T^2}\int \frac{d^3\textbf{\textit{p}}}{(2\pi)^3}\textbf{\textit{p}}^2 f_g^{eq}(\epsilon_1)\tilde{f}_g^{eq}(\epsilon_1)~.
	\end{align}
	From Eqs.~\eqref{zeta},~\eqref{apar}~and~\eqref{b}, we get
	\begin{align}\label{zetagluon1LL}
		\zeta _{gluon}^1&=\frac{1}{9T}\int \frac{d^3\textbf{\textit{p}}}{(2\pi)^3}\frac{\tau(\epsilon_1)}{\epsilon_1^2}f_g^{eq}\tilde{f}_g^{eq}\mathcal{A}_{par}^1 \nonumber \\
		& \times\bigg[\textbf{\textit{p}}^2-3c_s^2\epsilon_1^2+\frac{3c_s^2T}{2}\bigg(m_gm_g^\prime+\frac{m_g^3m_g^\prime-4\gamma_G^
			3\gamma_G^\prime}{\sqrt{m_g^4-4\gamma_G^4}}\bigg)\bigg]~,
	\end{align}
	where $f_g^{eq}\tilde{f}_g^{eq}=f_g^{eq}(\epsilon_1)\tilde{f}_g^{eq}(\epsilon_1)$.
	Doing the same calculation for $\epsilon=\epsilon_2$, we get
	\begin{align}\label{zetagluon2LL}
		\zeta _{gluon}^2&=\frac{1}{9T}\int \frac{d^3\textbf{\textit{p}}}{(2\pi)^3}\frac{\tau(\epsilon_2)}{\epsilon_2^2}f_g^{eq}\tilde{f}_g^{eq}\mathcal{A}_{par}^2\bigg[\textbf{\textit{p}}^2-3c_s^2\epsilon_2^2
		\nonumber \\
		& +\frac{3c_s^2T}{2}\bigg(m_gm_g^\prime+\frac{4\gamma_G^3\gamma_G^\prime-m_g^3m_g^\prime}{\sqrt{m_g^4-4\gamma_G^4}}\bigg)\bigg]~,
	\end{align}
	where $f_g^{eq}\tilde{f}_g^{eq}=f_g^{eq}(\epsilon_2)\tilde{f}_g^{eq}(\epsilon_2)$.


\end{document}